\documentclass[useAMS,usenatbib]{mnras}
\usepackage{natbib}
\usepackage{amsmath}
\usepackage{graphicx}
\usepackage{multirow}
\usepackage{url}
\usepackage[T1]{fontenc}
\usepackage{aecompl}
\numberwithin{equation}{section}
\title[Metallicity, radiation and chemistry in galaxies]{The effects of metallicity, UV radiation and non-equilibrium chemistry in high-resolution simulations of galaxies}
\author[A. J. Richings and J. Schaye]{A. J. Richings$^{1}$ and Joop Schaye$^{1}$\\
$^{1}$Leiden Observatory, Leiden University, PO Box 9513, 2300 RA Leiden, the Netherlands}

\begin{document}

\date{Accepted 2016 February 09. Received 2016 February 08 26; in original form 2015 June 28}

\pagerange{\pageref{firstpage}--\pageref{lastpage}} \pubyear{2016}

\maketitle

\label{firstpage}

\begin{abstract} 

We present a series of hydrodynamic simulations of isolated galaxies with stellar mass of $10^{9} \, \rm{M}_{\odot}$. The models use a resolution of $750 \, \rm{M}_{\odot}$ per particle and include a treatment for the full non-equilibrium chemical evolution of ions and molecules (157 species in total), along with gas cooling rates computed self-consistently using the non-equilibrium abundances. We compare these to simulations evolved using cooling rates calculated assuming chemical (including ionisation) equilibrium, and we consider a wide range of metallicities and UV radiation fields, including a local prescription for self-shielding by gas and dust. We find higher star formation rates and stronger outflows at higher metallicity and for weaker radiation fields, as gas can more easily cool to a cold (few hundred Kelvin) star forming phase under such conditions. Contrary to variations in the metallicity and the radiation field, non-equilibrium chemistry generally has no strong effect on the total star formation rates or outflow properties. However, it is important for modelling molecular outflows. For example, the mass of H$_{2}$ outflowing with velocities $> 50 \, \rm{km} \, \rm{s}^{-1}$ is enhanced by a factor $\sim 20$ in non-equilibrium. We also compute the observable line emission from C\textsc{ii} and CO. Both are stronger at higher metallicity, while C\textsc{ii} and CO emission are higher for stronger and weaker radiation fields respectively. We find that C\textsc{ii} is generally unaffected by non-equilibrium chemistry. However, emission from CO varies by a factor of $\sim 2 - 4$. This has implications for the mean $X_{\rm{CO}}$ conversion factor between CO emission and H$_{2}$ column density, which we find is lowered by up to a factor $\sim 2.3$ in non-equilibrium, and for the fraction of CO-dark molecular gas. 

\end{abstract}

\begin{keywords}
  astrochemistry - ISM: atoms - ISM: molecules - galaxies: evolution - galaxies: ISM. 
\end{keywords}

\section{Introduction}

Hydrodynamic simulations of galaxy formation typically model gas cooling by tabulating the cooling rate as a function of gas properties such as the density and temperature, under certain assumptions. For example, the simplest approach, as used in some of the first cosmological hydrodynamic simulations \citep[e.g.][]{katz92}, is to assume that the gas has primordial abundances and is in collisional ionisation equilibrium (CIE). \citet{sutherland93} also included the effects of metal-line cooling, computing cooling curves in CIE for a range of metallicities. 

Another effect that can be important for gas cooling is the presence of a photoionising UV radiation field, which can change the ionisation balance and heat the gas. \citet{efstathiou92} showed that an extragalactic UV background (UVB) can suppress the cooling rate in a primordial plasma, thereby inhibiting the formation of dwarf galaxies. \citet{katz96} implemented primordial radiative cooling in the presence of a UVB in cosmological hydrodynamic simulations. 

\citet{wiersma09} considered the impact that a photoionising UVB has on cooling rates in the presence of metals. They showed that photoionisation can suppress the cooling rate by up to an order of magnitude at temperatures and densities typical of the intergalactic medium (e.g. $10^{4} \, \rm{K} \la T \la 10^{6} \, \rm{K}$, $\rho / \left< \rho \right> \la 100$). They also showed that variations in relative abundances from their solar values can change the cooling rate by a factor of a few. \citet{wiersma09} tabulated the cooling rate from 11 elements separately in the presence of the redshift-dependent UVB of \citet{haardt01}, and these tables have been used in several cosmological hydrodynamic simulations \citep[e.g.][]{crain09, schaye10, schaye15, hopkins14}. 

The effects of metal cooling and UV radiation are particularly important below $10^{4} \, \rm{K}$, as cooling from atomic hydrogen becomes inefficient at such temperatures. In primordial gas, the only major coolants are H$_{2}$ and HD \citep[e.g.][]{saslaw67, flower00, glover08}. However, these molecules can easily be dissociated by Lyman-Werner radiation ($11.2 \, \rm{eV} - 13.6 \, \rm{eV}$). If metals are present, efficient cooling can continue down to $T \la 100 \, \rm{K}$ via fine-structure line emission \citep[e.g.][]{glover07,richings14a}. Conversely, the presence of UV radiation will hinder cooling to such low temperatures, as it heats the gas via photoionisation and photoelectric heating from dust grains, the latter of which also depends on metallicity. Metallicity and UV radiation therefore influence the transition from the warm ($\sim 10^{4} \, \rm{K}$) to the cold ($\sim 10^{2} \, \rm{K}$) gas phase. 

\citet{wolfire03} investigated the warm-to-cold transition by calculating the minimum pressure, and hence the minimum density, at which a cold phase can exist in pressure equilibrium with a warm phase, as a function of UV intensity, metallicity, dust abundance and ionisation rate from cosmic rays and Extreme UV (EUV)/X-ray radiation (see their equations 33-35). This minimum density increases with decreasing metallicity and increasing UV intensity (see also \citealt{glover14}). 

\citet{schaye04} demonstrated that the critical surface density above which a cold ISM phase can form in a galactic disc increases with decreasing metallicity and increasing UV intensity (see his equation 23). He additionally showed that the formation of a cold phase can trigger gravitational instabilities, due to the lower velocity dispersion in cold gas. Metallicity and UV radiation can thus affect the star forming properties of galaxies, in particular the threshold surface density below which star formation is unable to proceed. The transition from atomic to molecular hydrogen also depends on metallicity and UV radiation \citep[e.g.][]{schaye01, krumholz08, gnedin11, sternberg14}. 

Many of the above approaches assume that the gas is in chemical (including ionisation) equilibrium, or in other words, that the abundances of individual ions and molecules have reached an equilibrium or steady state. However, this assumption may not be valid if the dynamical or cooling time-scale of the gas is short compared to the chemical time-scale \citep[e.g.][]{kafatos73, sutherland93, gnat07, oppenheimer13a, vasiliev13}, or if the UV radiation field is varying on time-scales shorter than the chemical time-scale \citep[e.g.][]{oppenheimer13b}. 

Non-equilibrium chemistry has been shown to affect cooling rates in certain idealised scenarios. For example, in collisionally ionised gas that is cooling either isobarically or isochorically, the cooling rate computed self-consistently from non-equilibrium abundances is suppressed at temperatures $T > 10^{4} \, \rm{K}$ compared to gas in CIE \citep[e.g.][]{sutherland93, gnat07}. \citet{oppenheimer13a} investigated non-equilibrium ionisation and cooling in the presence of a photoionising radiation field, and they showed that the UVB reduces the non-equilibrium effects, although the cooling rates at $T > 10^{4} \, \rm{K}$ are still suppressed with respect to chemical equilibrium. \citet{richings14a} showed that, at temperatures below $10^{4} \, \rm{K}$, the direction of the non-equilibrium effects is reversed and the cooling rates are enhanced, due to an increase in the electron abundance compared to equilibrium. 

These studies considered special cases of a plasma that is cooling at constant density or pressure. \citet{walch11} compared equilibrium and non-equilibrium chemistry and cooling in high-resolution simulations of a turbulent medium, in a box $500 \, \rm{pc}$ across. They found that non-equilibrium chemistry did have a noticeable, albeit fairly small, effect on the gas temperature in their simulations. \citet{vasiliev13} applied a treatment for non-equilibrium cooling to simulations of a supernova remnant, and showed that the evolution of the supernova remnant is sensitive to non-equilibrium effects. However, it remains to be seen whether such non-equilibrium effects will be relevant on galactic scales. 

In this paper we investigate the effects of metallicity, UV radiation and non-equilibrium chemistry on galaxy formation. We run a series of hydrodynamic simulations of isolated galaxies with stellar and total halo masses of $10^{9}$ and $10^{11} \, \rm{M}_{\odot}$, respectively. The simulations use a resolution of $750 \, \rm{M}_{\odot}$ per particle and a gravitational force softening of $3.1 \, \rm{pc}$, which is sufficient to resolve the warm-to-cold transition. We consider a range of metallicities, $0.01 \, \rm{Z}_{\odot} \leq Z \leq \rm{Z}_{\odot}$, and a range of UV radiation fields that span nearly three orders of magnitude in H\textsc{i} photoionisation rate: the local interstellar radiation field (ISRF) of \citet{black87}, ten per cent of the \citet{black87} ISRF and the redshift zero extragalactic UV background (UVB) of \citet{haardt01}. We include a local prescription for self-shielding by gas and dust, and we also consider a model without self-shielding for comparison. We first run these simulations with the temperature and chemical abundances of the gas evolved using the non-equilibrium chemical model of \citet{richings14a, richings14b}, which follows the ionisation states of 11 elements that are important for the gas cooling rate, along with the abundances of 20 molecular species, including H$_{2}$ and CO. We then compare these to simulations evolved using cooling rates tabulated in chemical equilibrium. 

In addition to the dynamical impact on the galaxy due to the effects on the cooling rate, non-equilibrium chemistry can also affect observable diagnostics of individual chemical species, e.g. in the presence of a fluctuating UV field \citep{oppenheimer13b}, or in the presence of supersonic turbulence \citep{gray15}. To investigate how non-equilibrium chemistry can affect observable emission on galactic scales, we perform radiative transfer calculations in post-processing to compute line emission from C\textsc{ii} and CO. We then compare the intensity of line emission computed from non-equilibrium abundances to that computed assuming chemical equilibrium.

The remainder of this paper is organised as follows. In section~\ref{models_section} we describe the hydrodynamic methods and subgrid physics models used in our simulations. We present our simulations in section~\ref{simulations}, where we discuss the initial conditions (\ref{ic_section}), morphologies and star formation rates (\ref{sf_section}), outflow properties (\ref{outflows_section}), and the phase structure of the ISM (\ref{phase_structure}). We compute the observable line emission from C\textsc{ii} and CO in section~\ref{line_emission}, and we summarise our main results in section~\ref{conclusions}. 

\section{Subgrid Physics Models}\label{models_section}

Our simulations were run using a modified version of the tree/Smoothed Particle Hydrodynamics (SPH) code \textsc{gadget}3 \citep[last described in][]{springel05}. The hydrodynamic equations were solved using the set of numerical methods known as \textsc{anarchy}, which includes many of the latest improvements to the standard SPH implementation. In particular, \textsc{anarchy} uses the pressure-entropy formulation of SPH, derived by \citet{hopkins13}; the artificial viscosity switch of \citet{cullen10}; a switch for artificial conduction, similar to that used in \citet{price08}; the time-step limiters of \citet{durier12}; and the $C^{2}$ \citet{wendland95} kernel. \textsc{anarchy} will be described in more detail in Dalla Vecchia (in preparation); see also Appendix A of \citet{schaye15} for a description of the implementation of \textsc{anarchy} in the cosmological simulations that were run for the \textsc{eagle} project, which is identical to the \textsc{anarchy} implementation that we use here. 

We use prescriptions to model physical processes that are unresolved in our simulations, including the chemical evolution of ions and molecules, radiative cooling, star formation and stellar feedback. These subgrid models are summarised below. 

\subsection{Chemistry and radiative cooling}\label{chemical_model}

We use the chemical model of \citet{richings14a, richings14b} to evolve the chemical abundances of 157 species, including all ionisation states of the 11 elements that are most important for cooling\footnote{H, He, C, N, O, Ne, Mg, Si, S, Ca \& Fe.} and 20 molecular species\footnote{H$_{2}$, H$_{2}^{+}$, H$_{3}^{+}$, OH, H${_2}$O, C$_{2}$, O$_{2}$, HCO$^{+}$, CH, CH$_{2}$, CH$_{3}^{+}$, CO, CH$^{+}$, CH$_{2}^{+}$, OH$^{+}$, H$_{2}$O$^{+}$, H$_{3}$O$^{+}$, CO$^{+}$, HOC$^{+}$, O$_{2}^{+}$.}. The gas temperature is evolved using radiative cooling and heating rates calculated from the non-equilibrium abundances. This gives us a set of 158 differential equations (157 chemical rate equations and the energy equation), which we integrate over each hydrodynamic timestep for each gas particle. We integrate these differential equations using the backward difference formula method and Newton iteration in \textsc{cvode} (from the \textsc{sundials}\footnote{\url{https://computation.llnl.gov/casc/sundials/main.html}} suite of non-linear differential/algebraic equation solvers), using a relative tolerance of $10^{-4}$ and an absolute tolerance of $10^{-10}$. 

We summarise the main chemical and thermal processes that are included in our model below. 

\subsubsection{Chemical processes}\label{chemical_processes}

We include collisional ionisation, radiative and di-electronic recombinations and charge transfer reactions. The formation of molecular hydrogen occurs on dust grains, using equation 18 of \citet{cazaux02} with a dust temperature $T_{\rm{dust}} = 10 \, \rm{K}$, as well as via gas phase reactions. We also include cosmic ray ionisations, assuming a primary ionisation rate of atomic hydrogen due to cosmic rays of $\zeta_{\rm{HI}} = 2.5 \times 10^{-17} \text{s}^{-1}$ \citep{williams98}. The primary ionisation rates of other species due to cosmic rays are then scaled to this value using the ratios in the \textsc{umist} database\footnote{\url{http://www.udfa.net/}} \citep{mcelroy13} where available, or using the equations from \citet{lotz67}, \citet{silk70} and \citet{langer78} otherwise. We include secondary ionisations of H\textsc{i} and He\textsc{i} from cosmic rays using the tables of \citet{furlanetto10}, assuming a typical mean primary electron energy of $E = 35 \, \rm{eV}$. 

\begin{table}
\centering
\begin{minipage}{84mm}
\caption{Properties of the UV radiation fields considered in this paper.}
\centering
\begin{tabular}{lcccc}
\hline
UV field & $G_{0}$\footnote{Radiation field strength in the energy range $6.0 - 13.6 \, \rm{eV}$, in units of the \citet{habing68} field.} & $\Gamma_{\rm{H\textsc{i}}} \, (\rm{s}^{-1})$\footnote{Unattenuated H\textsc{i} photoionisation rate.} & $\Gamma_{\rm{He\textsc{i}}} \, (\rm{s}^{-1})$\footnote{Unattenuated He\textsc{i} photoionisation rate.} & $\Gamma_{\rm{He\textsc{ii}}} \, (\rm{s}^{-1})$\footnote{Unattenuated He\textsc{ii} photoionisation rate.} \\
\hline
ISRF & $1.2$ & $4.4 \times 10^{-11}$ & $3.7 \times 10^{-12}$ & $1.7 \times 10^{-14}$ \\ 
lowISRF & $0.12$ & $4.4 \times 10^{-12}$ & $3.7 \times 10^{-13}$ & $1.7 \times 10^{-15}$ \\ 
UVB & $0.014$ & $8.4 \times 10^{-14}$ & $2.0 \times 10^{-14}$ & $1.5 \times 10^{-16}$ \\ 
\hline
\vspace{-0.2in}
\label{uv_table}
\end{tabular}
\end{minipage}
\end{table}

We include photoionisation of atoms and ions, including Auger ionisation, using optically thin cross sections calculated in the grey approximation for a given UV spectrum. We consider three UV spectra in this paper: the local ISRF of \citet{black87}, ten per cent of the \citet{black87} ISRF and the redshift zero extragalactic UVB of \citet{haardt01}. The properties of these radiation fields are summarised in Table~\ref{uv_table}. We then attenuate the optically thin rates by the gas and dust, as a function of the column densities of H\textsc{i}, H$_{2}$, He\textsc{i}, He\textsc{ii} and dust, as described in \citet{richings14b}. To calculate these column densities, we assume that each gas particle is shielded locally. The total hydrogen column density, $N_{\rm{H_{tot}}}$, is then: 

\begin{equation}\label{column_equation}
N_{\rm{H_{tot}}} = n_{\rm{H_{tot}}} L \, , 
\end{equation}
where $n_{\rm{H_{tot}}}$ is the hydrogen number density of the gas particle and $L$ is the shielding length, i.e. the length scale over which the gas particle is able to shield itself. We use a Sobolev-like approximation to estimate the shielding length based on the gradient of the density, $\rho$: 

\begin{equation}\label{sobolev_length}
L = L_{\rm{Sob}, \rho} = \frac{\rho}{\lvert 2 \nabla \rho \rvert}.
\end{equation}
\citet{gnedin09} use this approximation in cosmological simulations to follow the formation of molecular hydrogen. They show that this approximation accurately reproduces the true column densities in their simulations, as measured along random lines of sight, with a scatter of a factor of 2 in the range $3 \times 10^{20} \, \rm{cm}^{-2} < N_{\rm{HI}} + 2 N_{\rm{H_{2}}} < 3 \times 10^{23} \, \rm{cm}^{-2}$. 

The column density of species $i$ is then: 

\begin{equation}
N_{i} = x_{i} N_{\rm{H_{tot}}}, 
\end{equation}
where $x_{i} = n_{i} / n_{\rm{H_{tot}}}$ is the abundance of species $i$. 

The photodissociation rates of molecular species are also attenuated using the shielding length given in equation~\ref{sobolev_length}. The attenuation of the photodissociation rate of species $i$ due to dust is given by: 

\begin{equation} 
S_{\rm{d}}^{i}(N_{\rm{H_{tot}}}, Z) = \exp(- \gamma_{d}^{i} A_{\rm{v}}), 
\end{equation}
where $S_{\rm{d}}^{i}(N_{\rm{H_{tot}}}, Z)$ is the shielding factor (i.e. the ratio of the optically thick to optically thin photodissociation rates) due to dust, $A_{\rm{v}} = 4.0 \times 10^{-22} N_{\rm{H_{tot}}} Z / Z_{\odot} \, \rm{mag}$ is the visual dust extinction, $Z$ is the metallicity and the factors $\gamma_{\rm{d}}^{i}$ are taken from table 2 of \citet{vandishoeck06} where available, or from table B2 of \citet{glover10} otherwise. 

In addition to dust shielding, molecular hydrogen can also be self-shielded. We use the temperature-dependent self-shielding function given in equations 3.12 to 3.15 of \citet{richings14b}, which gives the shielding factor of H$_{2}$ as a function of the molecular hydrogen column density, $N_{\rm{H_{2}}}$, gas temperature, $T$, and Doppler broadening parameter, $b$. The Doppler broadening parameter includes thermal broadening and broadening due to turbulence. In our simulations, we assume a constant turbulent broadening parameter $b_{\rm{turb}} = 7.1 \, \rm{km} \, \rm{s}^{-1}$, as used by e.g. \citet{krumholz12}, as this is typical for observed molecular clouds. 

CO can also be self-shielded and shielded by H$_{2}$. We use the tables of the shielding factor of CO as a function of CO and H$_{2}$ column densities given by \citet{visser09}. 

A full list of the chemical reactions that are included in our model can be found in table B1 of \citet{richings14a}. 

\subsubsection{Thermal processes}

We use the non-equilibrium abundances from the chemical network to calculate the net cooling rate of the gas. We include cooling from the collisional excitation and ionisation of H and He, fine-structure line emission from metals, recombination cooling and bremsstrahlung radiation. We use the H$_{2}$ cooling function from \citet{glover08} for rovibrational cooling from H$_{2}$, and we include cooling from CO, H$_{2}$O and OH. The photoheating rate is attenuated by gas and dust, as described by \citet{richings14b}, using column densities calculated from equations~\ref{column_equation} and \ref{sobolev_length} above. We also include photoelectric heating on dust grains \citep{bakes94, wolfire95}, cosmic ray heating \citep{goldsmith78, glover07}, and heating from the photodissociation of H$_{2}$ \citep{black77}, UV pumping of H$_{2}$ \citep{burton90}, and the formation of H$_{2}$ on dust grains \citep{hollenbach79} and in the gas phase \citep{karpas79, launay91}. 

\subsubsection{Equilibrium cooling tables}\label{cooling_tables}

One aim of this paper is to compare simulations run using the full non-equilibrium chemical model described above to simulations evolved with cooling rates that are calculated assuming chemical equilibrium. We therefore used our chemical model to create tables of the cooling and heating rates and the mean molecular weight in chemical equilibrium as a function of gas temperature, hydrogen number density and total hydrogen column density. Note that the simulations in this paper were run at fixed metallicity, so we do not need to include an additional dimension for metallicity in the cooling tables. These tables were then used to evolve the gas temperature in the equilibrium cooling runs. By computing equilibrium cooling rates using the same chemical model as used for the non-equilibrium runs, we ensure that any differences are due to non-equilibrium effects, and not simply due to the use of different chemical rates in the models. 

\subsection{Star formation}

We allow a gas particle to form stars if its temperature, $T$, and hydrogen number density, $n_{\rm{H_{tot}}}$, satisfy the following criteria: 

\begin{align}
T &< T_{\rm{thresh}} = 1000 \, \rm{K}, \\
n_{\rm{H_{tot}}} &> n_{\rm{H_{tot}, thresh}} = 1.0 \, \rm{cm}^{-3}. 
\end{align}

Gas particles that satisfy these criteria form stars at a rate given by: 

\begin{equation}\label{sfr_eqn}
\dot{\rho}_{\ast} = \epsilon_{\rm{SF}} \frac{\rho_{\rm{gas}}}{t_{\rm{ff}}}, 
\end{equation}
where $\dot{\rho}_{\ast}$ is the star formation rate per unit volume, $\rho_{\rm{gas}}$ is the gas density and $t_{\rm{ff}} = \sqrt{3 \pi / (32 G \rho)}$ is the free fall time. We use a star formation efficiency per free fall time of $\epsilon_{\rm{SF}} = 0.005$. This is slightly lower than the value of $\epsilon_{\rm{SF}} \sim 0.015$ that is typically observed in the Milky Way and nearby galaxies \citep{krumholzetal12}. We calibrated this parameter, along with the density and temperature thresholds and the parameters for the stellar feedback model described in the next section, to reproduce the observed Kennicutt-Schmidt relation in nearby galaxies (see Fig.~\ref{KS}). 

In a timestep $\Delta t$, star forming gas particles are stochastically turned into star particles with a probability $p$ given by: 

\begin{equation}
p = \min \left( \frac{\dot{\rho}_{\ast} \Delta t}{\rho_{\rm{gas}}}, \, 1 \right). 
\end{equation}

\subsection{Stellar feedback}\label{feedback_section}

Each star particle represents a stellar population, rather than an individual star. We assume that the stellar population initially follows the \citet{chabrier03} Initial Mass Function (IMF) with masses in the range $0.1 - 100 \, \rm{M}_{\odot}$. We then calculate how many type II supernovae will explode in each timestep for each star particle, using the metallicity-dependent stellar lifetimes of \citet{portinari98} and assuming that all stars with a mass greater than $6 \, \rm{M}_{\odot}$ will end their lives in a supernova (stars with masses of $6 - 8 \, \rm{M}_{\odot}$ explode as electron capture supernovae in models with convective overshoot, e.g. \citealt{chiosi92}). 

The energetic feedback from supernovae is implemented using the stochastic thermal feedback prescription of \citet{dallavecchia12}, except that we distribute the total available energy from type II supernovae from a single star particle in time according to the stellar lifetimes of the massive stars, rather than combine it into a single supernova event. For each star particle that has a non-zero number of supernovae, $N_{\rm{SNII}}$, in timestep $\Delta t$, we stochastically select gas particles from the $N_{\rm{ngb}} = 48$ neighbours to be heated by $\Delta T = 10^{7.5} \, \rm{K}$. Note that $N_{\rm{ngb}}$ is smaller than the number of neighbours that we use to compute the SPH kernel ($N_{\rm{ngb}}^{\rm{SPH}} = 100$). This reduces the computational cost of the neighbour-finding routine for star particles, and has little impact on the accuracy of the stellar feedback. The probability $p$ of selecting a given gas particle to be heated is: 

\begin{equation}
p = \min \left( \frac{2}{3} \frac{E_{\rm{SNII}} N_{\rm{SNII}} \mu m_{\rm{p}}}{k_{\rm{B}} \Delta T \sum_{i = 1}^{N_{\rm{ngb}}} m_{i}}, \, 1 \right), 
\end{equation}
where $E_{\rm{SNII}} = 10^{51} \, \rm{erg}$ is the energy injected per supernova, $\mu$ is the mean molecular mass of the gas particle, $m_{\rm{p}}$ is the mass of a proton and $m_{i}$ is the mass of the $i^{\rm{th}}$ neighbouring gas particle. 

By imposing a minimum heating temperature $\Delta T$, we ensure that the cooling time of the heated gas is long enough to avoid artificial radiative losses that would otherwise make the feedback scheme ineffective. However, note that, at the resolution that we use in our simulations ($750 \, \rm{M}_{\odot}$ per particle), we require approximately 10 supernovae to heat a single gas particle to $10^{7.5} \, \rm{K}$. We therefore do not resolve individual supernovae, as each feedback event is equivalent to several supernovae exploding simultaneously. 

While we include the energetic feedback from supernovae, we do not include the chemical enrichment from supernovae or from stellar mass loss. This allows us to follow model galaxies at a fixed metallicity, which eases the interpretation of our numerical experiments. 

\subsection{Jeans limiter}

To ensure that we always resolve the Jeans mass, $M_{\rm{J}}$, and the Jeans length, $L_{\rm{J}}$, we impose a minimum floor on the pressure that enters the hydrodynamic equations. This is similar to the approach used by e.g. \citet{robertson08, schaye08, hopkins11}. 

There are two criteria that we can consider to determine the pressure floor. Firstly, we can require that the Jeans mass be resolved by at least a factor $N_{\rm{J, \, m}}$ times the mass within the SPH kernel. The Jeans mass is given by: 

\begin{equation}\label{jeans_mass}
M_{\rm{J}} = \frac{\pi^{5 / 2} c_{\rm{s}}^{3}}{6 G^{3 / 2} \rho^{1 / 2}}, 
\end{equation}
where $c_{\rm{s}} = \sqrt{\gamma P / \rho}$ is the sound speed, $\gamma = 5 / 3$ is the ratio of specific heats, $P$ is the thermal gas pressure and $\rho$ is the gas density. 

If each gas particle has mass $m_{\rm{gas}}$, and we use $N_{\rm{ngb}}^{\rm{SPH}}$ neighbours in the SPH kernel, then the mass within the kernel is: 

\begin{equation}\label{kernel_mass}
M_{\rm{k}} = N_{\rm{ngb}}^{\rm{SPH}} m_{\rm{gas}}. 
\end{equation}
From equations~\ref{jeans_mass} and \ref{kernel_mass}, we find that the minimum pressure that we require is:

\begin{equation}\label{floor_mass}
P_{\rm{floor, \, m}} = \left( \frac{36}{\pi^{5}} \right)^{1/3} \frac{G}{\gamma} (N_{\rm{J, \, m}} N_{\rm{ngb}}^{\rm{SPH}} m_{\rm{gas}})^{2/3} \rho^{4/3}. 
\end{equation}

Secondly, we can require that the Jeans length be resolved by at least a factor $N_{\rm{J, \, l}}$ times the SPH smoothing length, $h_{\rm{sml}}$. Using equation~\ref{jeans_mass}, the Jeans length is: 

\begin{equation}
L_{\rm{J}} = \frac{\pi^{1/2} c_{\rm{s}}}{G^{1/2} \rho^{1/2}}. 
\end{equation}
Hence the minimum pressure that we require is: 

\begin{equation}\label{floor_length}
P_{\rm{floor, \, l}} = \frac{G}{\pi \gamma} N_{\rm{J, \, l}}^{2} h_{\rm{sml}}^{2} \rho^{2}. 
\end{equation}

Since the smoothing length is defined such that the number of SPH neighbours, $N_{\rm{ngb}}^{\rm{SPH}}$, is constant, we can express $h_{\rm{sml}}$ in terms of the kernel mass as: 

\begin{equation}\label{sml_equation}
h_{\rm{sml}} = \left( \frac{3 N_{\rm{ngb}}^{\rm{SPH}} m_{\rm{gas}}}{4 \pi \rho} \right)^{1/3}. 
\end{equation}
By combining equations~\ref{floor_mass}, \ref{floor_length} and \ref{sml_equation}, we see that these two criteria are equivalent if: 

\begin{equation}
N_{\rm{J, \, l}} = 2 N_{\rm{J, m}}^{1/3}. 
\end{equation}

We set the pressure floor in our simulations such that we resolve $M_{\rm{J}}$ by at least $N_{\rm{J, \, m}} = 4$ kernel masses, and thus we resolve $L_{\rm{J}}$ by at least $N_{\rm{J, \, l}} = 3.2$ smoothing lengths. 

Note that, while we impose this floor on the pressure that enters the hydrodynamic equations, the temperature is still allowed to cool below this limit. This means that the thermal pressure will effectively be decoupled from the hydrodynamic equations once this floor is reached. For the resolution of our simulations, the pressure floor that we use corresponds to the thermal pressure at a temperature of $240 \, \rm{K}$ at the star formation threshold density $n_{\rm{H_{tot}, thresh}} = 1.0 \, \rm{cm}^{-3}$. We implement the Jeans limiter as a pressure floor rather than a temperature floor (as was used in \citealt{schaye08}) so that the thermal and chemical state of the gas will evolve towards a realistic equilibrium for the given density, although we miss small-scale, high-density structures that are unresolved in the simulation. 

\section{Simulations}\label{simulations}

We ran a suite of SPH simulations of isolated galaxies with a range of metallicities and UV radiation fields, using a resolution of $750 \, \rm{M}_{\odot}$ per gas particle, with $100$ SPH neighbours, and a gravitational softening length of $3.1 \, \rm{pc}$. Each simulation was evolved for 1 Gyr. We ran each simulation twice: once using the full non-equilibrium chemical model summarised in section~\ref{chemical_model}, and once using tabulated cooling rates calculated assuming chemical equilibrium, as described in section~\ref{cooling_tables}. In this section, we describe the initial conditions, and we compare the morphologies, star formation rates, outflow properties and ISM phases in our simulated galaxies. For each of these, we focus on the effects of metallicity, radiation field and non-equilibrium chemistry. 

\subsection{Initial conditions}\label{ic_section}

\begin{table*}
\centering
\begin{minipage}{168mm}
\caption{Parameters and properties of the galaxies in the suite of simulations used in this paper: total mass $M_{200}$ within the radius $R_{200, \rm{crit}}$ enclosing a mean density of 200 times the critical density of the Universe at redshift zero, concentration $c_{200}$ of the dark matter halo, initial stellar mass $M_{\ast, \, \rm{init}}$, initial gas mass $M_{\rm{gas, \, init}}$, disc gas mass fraction $f_{\rm{d, \, gas}}$, mass per gas or star particle $m_{\rm{baryon}}$, gravitational softening length $\epsilon_{\rm{soft}}$, gas metallicity $Z$, UV radiation field and whether we include self-shielding.}
\centering
\begin{tabular}{lcccccccccc}
\hline
Model & $M_{200}$ & $c_{200}$ & $M_{\ast, \, \rm{init}}$ & $M_{\rm{gas, \, init}}$ & $f_{\rm{d, \, gas}}$ & $m_{\rm{baryon}}$ & $\epsilon_{\rm{soft}}$ & $Z$ & UV Field\footnote{See Table~\ref{uv_table}} & Shielding \\
 & ($\rm{M}_{\odot}$) & & ($\rm{M}_{\odot}$) & ($\rm{M}_{\odot}$) & & ($\rm{M}_{\odot}$) & ($\rm{pc}$) & ($\rm{Z}_{\odot}$) & & \\
\hline
ref & $10^{11}$ & 8.0 & $1.4 \times 10^{9}$ & $4.8 \times 10^{8}$ & 0.3 & 750 & 3.1 & 0.1 & ISRF & yes \\
lowZ & $10^{11}$ & 8.0 & $1.4 \times 10^{9}$ & $4.8 \times 10^{8}$ & 0.3 & 750 & 3.1 & \textbf{0.01} & ISRF & yes \\
hiZ & $10^{11}$ & 8.0 & $1.4 \times 10^{9}$ & $4.8 \times 10^{8}$ & 0.3 & 750 & 3.1 & \textbf{1.0} & ISRF & yes \\
lowISRF & $10^{11}$ & 8.0 & $1.4 \times 10^{9}$ & $4.8 \times 10^{8}$ & 0.3 & 750 & 3.1 & 0.1 & \textbf{lowISRF} & yes \\
UVB & $10^{11}$ & 8.0 & $1.4 \times 10^{9}$ & $4.8 \times 10^{8}$ & 0.3 & 750 & 3.1 & 0.1 & \textbf{UVB} & yes \\
UVBthin & $10^{11}$ & 8.0 & $1.4 \times 10^{9}$ & $4.8 \times 10^{8}$ & 0.3 & 750 & 3.1 & 0.1 & \textbf{UVB} & \textbf{no} \\
\hline
\vspace{-0.3in}
\label{sim_parameters}
\end{tabular}
\end{minipage}
\end{table*}

The initial conditions that we use are based on the model of \citet{springeletal05}, and were generated using a modified version of a code that was kindly provided to us by Volker Springel. Each galaxy consists of an exponential disc of gas and stars with a radial scale length of $2.0 \, \rm{kpc}$, and a central stellar bulge with a \citet{hernquist90} density profile, embedded within a dark matter halo. The stellar disc and bulge are represented by collisionless particles with the same mass as the gas particles ($750 \, \rm{M}_{\odot}$ per particle). The main difference between our initial conditions and the model of \citet{springeletal05} is that we represent the dark matter halo using a static potential, rather than with live dark matter particles. Using a static potential speeds up the calculation without affecting the results. 

The gas is initially isothermal, with a temperature of $10^{4} \, \rm{K}$, and the chemical abundances are initially in chemical equilibrium. The stellar disc is set up with a vertical distribution that follows an isothermal profile with a scale height of ten per cent of the radial scale length of the disc, while the vertical structure of the gaseous disc is set to be in hydrostatic equilibrium using an iterative procedure. 

The total mass of each galaxy within $R_{200, \rm{crit}}$ (i.e. the radius enclosing a sphere with a mean density of 200 times the critical density of the Universe at redshift zero) is $M_{200} = 10^{11} \, \rm{M}_{\odot}$, and the initial stellar mass is $M_{\ast} = 1.4 \times 10^{9} \, \rm{M}_{\odot}$. These masses are consistent with the stellar mass-halo mass relation obtained from abundance matching by \citet{moster13} and corrected for baryonic effects using the prescription of \citet{sawala15}. 

A fraction $f_{\ast, \rm{B}} = 0.2$ of the stellar mass is in the bulge, with the remainder in the disc. We use a disc gas mass fraction $f_{\rm{d, \, gas}} = 0.3$, which gives an initial gas mass of $M_{\rm{gas}} = 1.8 \times 10^{8} \, \rm{M}_{\odot}$. The dark matter density profile that we use to calculate the static dark matter potential follows a \citet{hernquist90} profile that has been scaled to match the inner regions of a \citet{navarro96} (NFW) profile with a concentration $c_{200} = 8.0$, which agrees with the redshift zero mass-concentration relation of \citet{duffy08}. 

In our reference model (ref), we use a fixed metallicity of $0.1 \, \rm{Z}_{\odot}$, and for the UV radiation field we use the ISRF of \citet{black87}, along with the self-shielding prescription described in section~\ref{chemical_processes}. We also consider two additional metallicities ($0.01 \, \rm{Z}_{\odot}$ and $\rm{Z}_{\odot}$), and three additional radiation fields (ten per cent of the \citet{black87} ISRF and the redshift zero UVB of \citet{haardt01}, both with self-shielding, and the redshift zero \citet{haardt01} UVB without self-shielding). The parameters and properties of the galaxies in our suite of simulations are summarised in Table~\ref{sim_parameters}. We summarise the properties of the different radiation fields used in our simulations in Table~\ref{uv_table}. 

Throughout this paper we assume that the dust-to-gas ratio scales linearly with metallicity, or in other words, that the dust-to-metals ratio is constant. However, there is observational evidence that the dust-to-metals ratio decreases at low metallicity, below $\approx 0.3 \, \rm{Z}_{\odot}$ \citep[e.g.][]{remyruyer14}. Therefore, it is possible that our runs at $0.01 \, \rm{Z}_{\odot}$ and $0.1 \, \rm{Z}_{\odot}$ overestimate the total dust abundance. This would affect the formation rate of H$_{2}$, shielding of the radiation field, and photoelectric heating from dust grains. 

\subsection{Morphology and star formation}\label{sf_section}

\begin{figure*}
\centering
\mbox{
	\includegraphics[width=112mm]{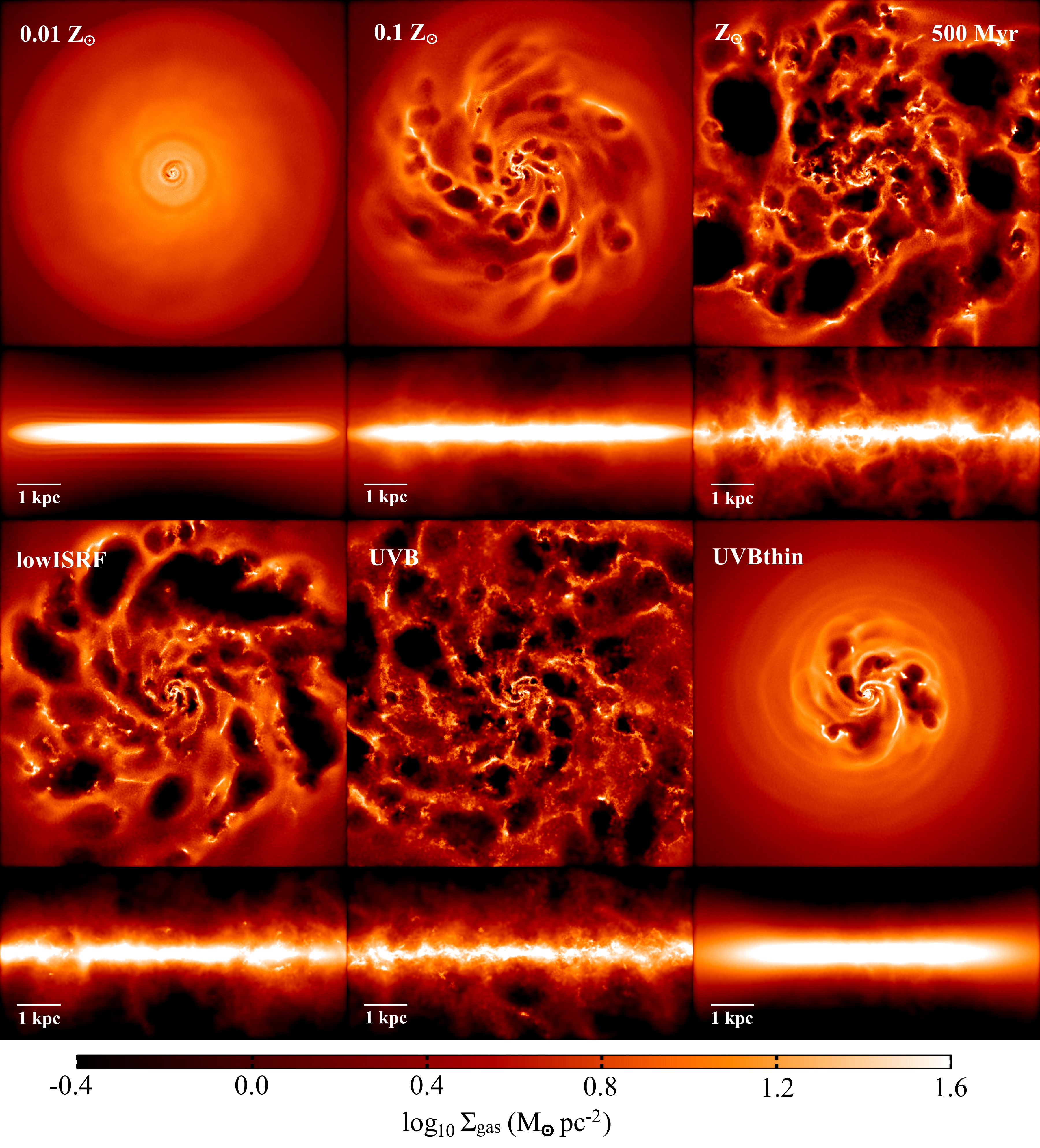}}
\caption{Maps of the total gas surface density after 500 Myr in the simulations run with the full non-equilibrium chemical model. The simulations in the top row were run with the \citet{black87} ISRF at different metallicities: $0.01 \, \rm{Z}_{\odot}$ \textit{(lowZ; left)}, $0.1 \, \rm{Z}_{\odot}$ \textit{(ref; centre)} and $\rm{Z}_{\odot}$ \textit{(hiZ; right)}. The simulations in the bottom row were run at a metallicity of $0.1 \, \rm{Z}_{\odot}$ with different UV radiation fields: ten per cent of the \citet{black87} ISRF \textit{(lowISRF; left)}, the \citet{haardt01} UVB at redshift zero \textit{(UVB; centre)} and the \citet{haardt01} UVB without self-shielding \textit{(UVBthin; right)}. Each pair of panels shows projections looking at the disc face-on \textit{(top)} and edge-on \textit{(bottom)}. The face-on projections are $8 \, \rm{kpc}$ across, and the edge-on projections cover $4 \, \rm{kpc}$ in the vertical direction.}
\label{gasDensityFig}
\end{figure*}

\begin{figure*}
\centering
\mbox{
	\includegraphics[width=112mm]{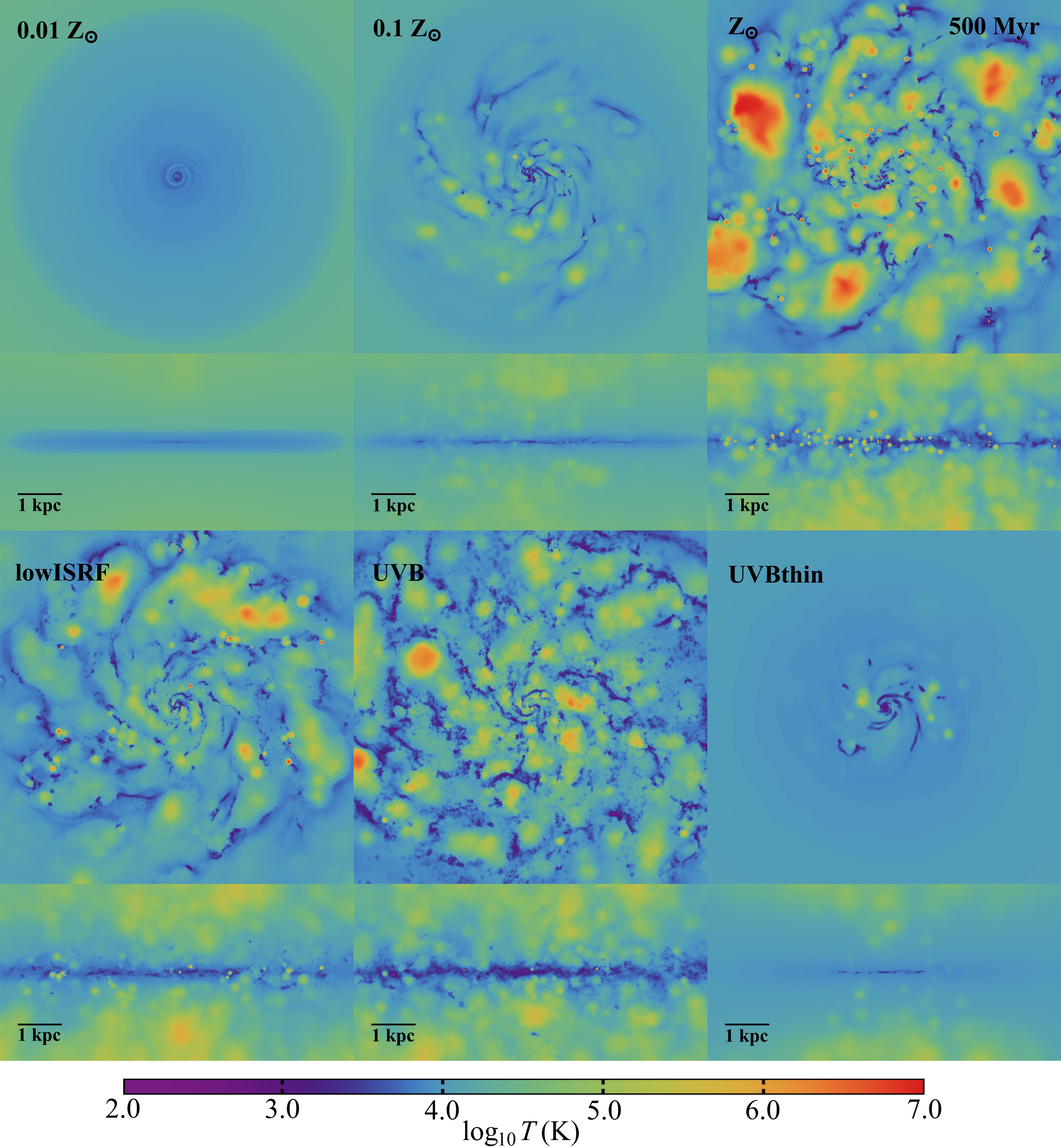}}
\caption{As Fig.~\ref{gasDensityFig}, but for the gas temperature. Except for $Z = 0.01 \, \rm{Z}_{\odot}$, hot bubbles of gas driven by supernovae disrupt the otherwise smooth density distribution in the disc and drive vertical outflows of gas.}
\label{gasTemperatureFig}
\end{figure*}

Maps of the gas surface density after $500 \, \rm{Myr}$ from the simulations run with the full non-equilibrium chemical model are shown in Fig.~\ref{gasDensityFig} for different metallicities (top row) and different UV radiation fields (bottom row), and maps of the gas temperature after $500 \, \rm{Myr}$ are shown in Fig.~\ref{gasTemperatureFig}. Each pair of panels in these figures shows projections looking at the disc face-on (top) and edge-on (bottom). 

Comparing the three different metallicities, we see that the morphology of the gas is very different in these three runs. In the simulation with the lowest metallicity (one per cent solar; top left pair of panels in Figs.~\ref{gasDensityFig} and \ref{gasTemperatureFig}), star formation only occurs at the centre of the disc. However, at higher metallicities, star formation becomes more vigorous and extends to larger radii. This leads to more supernovae, which create more bubbles of hot gas in the disc (as seen in Fig.~\ref{gasTemperatureFig}), and drive more gas out of the disc in vertical fountains and outflows (as seen in the edge-on views). 

\begin{figure*}
\centering
\mbox{
	\includegraphics[width=168mm]{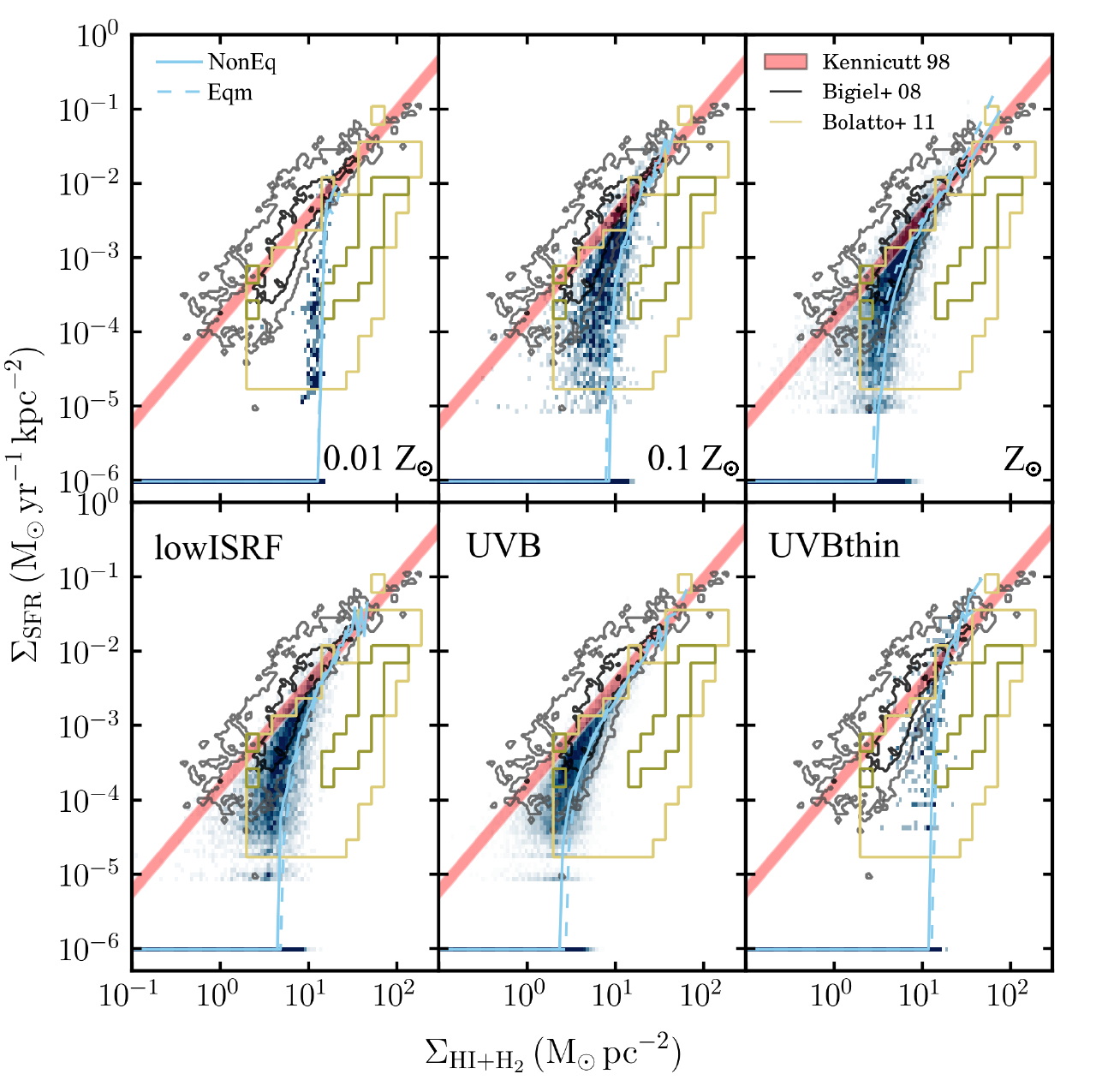}}
\caption{The star formation rate surface density, $\Sigma_{\rm{SFR}}$, versus surface density of neutral (atomic plus molecular) hydrogen, $\Sigma_{\rm{HI + H_{2}}}$, i.e. the Kennicutt-Schmidt relation. The intensity of the blue-green colour corresponds to two-dimensional histograms of $\Sigma_{\rm{SFR}}$ and $\Sigma_{\rm{HI + H_{2}}}$ measured in the simulations run with the full non-equilibrium chemical model at different metallicities \textit{(top row)} and for different radiation fields \textit{(bottom row; see Table~\ref{uv_table})}. We measure $\Sigma_{\rm{SFR}}$ and $\Sigma_{\rm{HI + H_{2}}}$ on a grid of cells, each $100 \, \rm{pc}$ across, that span a square region $8 \, \rm{kpc}$ across aligned such that the disc is viewed face on. We combine measurements from all snapshot outputs from $100 \, \rm{Myr}$ to $1000 \, \rm{Myr}$, at intervals of $100 \, \rm{Myr}$. We set all cells to have a minimum $\Sigma_{\rm{SFR}}$ of $10^{-6} \, \rm{M}_{\odot} \, \rm{yr}^{-1} \, \rm{kpc}^{-2}$, so that they are visible in these plots. The cyan curves show the median value of $\Sigma_{\rm{SFR}}$ in bins of $\Sigma_{\rm{HI + H_{2}}}$, from simulations evolved with the full non-equilibrium chemical model \textit{(solid)} and those run with cooling rates in chemical equilibrium \textit{(dashed)}. The contours show the observed Kennicutt-Schmidt relation from various samples. The red shaded region shows the best-fit power-law relation of \citet{kennicutt98} (their equation 4), with the width of this region indicating the uncertainty in the normalisation. The black and grey contours are taken from \citet{bigiel08} (the centre-right panel of their fig. 8), measured in sub-kpc regions of nearby galaxies. Finally, the yellow contours are take from \citet{bolatto11} (from their fig. 6), measured in the Small Magellanic Cloud. We use their data at a resolution of $200 \, \rm{pc}$, rather than their full resolution data, as this is closer to the spatial scale that we use for the measurements of $\Sigma_{\rm{SFR}}$ and $\Sigma_{\rm{HI + H_{2}}}$ in the simulations ($100 \, \rm{pc}$). We have renormalised $\Sigma_{\rm{SFR}}$ from the observations for a \citet{chabrier03} IMF, as used in our simulations.}
\label{KS}
\end{figure*}

These trends with metallicity occur because, at higher metallicities, there is more metal-line cooling, which allows the gas to cool down to a cold ISM phase (with temperatures of a few hundred Kelvin) at lower densities. Gas needs to cool down to the cold ISM phase before it can form stars. Therefore, in the runs at higher metallicity, star formation can proceed in lower-density gas, and hence at lower gas surface densities. Our models all start with the same gas density profile. Hence, if star formation extends to lower gas surface densities, then it will extend to larger radii. 

Additionally, the increase in metal-line cooling at higher metallicities means that gas can more easily undergo gravitational collapse to higher densities. These two effects (star formation at lower densities and more gas collapsing to higher densities) lead to an increase in the total star formation rate, and hence more stellar feedback in the disc. 

Furthermore, the fact that the Jeans scale is smaller at lower temperatures also contributes to the more fragmented appearance of the gas disc at high metallicities. 

However, it is important to note that these runs at different metallicities all use the same radiation field. In reality, the lower star formation rate that we find at lower metallicity will also result in a weaker radiation field. As we show below, this will tend to increase the star formation rate and thus will lessen the differences in star formation rate at different metallicities. 

The bottom rows of Figs.~\ref{gasDensityFig} and \ref{gasTemperatureFig} show runs at ten per cent solar metallicity with different UV radiation fields: ten per cent of the \citet{black87} ISRF (`lowISRF') and the redshift zero extragalactic UVB of \citet{haardt01} (`UVB'), both run with self-shielding, and the redshift zero \citet{haardt01} UVB without self-shielding (`UVBthin'). See Table~\ref{uv_table} for the properties of these radiation fields. As we decrease the strength of the UV radiation field (from the reference run in the top centre panels of Figs.~\ref{gasDensityFig} and \ref{gasTemperatureFig}, to `lowISRF' to `UVB'), the gas disc becomes more fragmented. A lower radiation field reduces the heating rate from photoionisation and photoelectric dust heating, which allows the gas to cool to the cold, star forming ISM phase (with temperatures of a few hundred Kelvin) at lower densities. This increases the star formation rate, leading to more stellar feedback, and it decreases the Jeans scale in this gas. Hence the gas becomes more fragmented, similar to the effect of increasing the metallicity. 

In the run without self-shielding (`UVBthin'; bottom right pair of panels in Figs.~\ref{gasDensityFig} and \ref{gasTemperatureFig}), the gas is heated by photoionisation of H\textsc{i} even at high densities, where it would otherwise become shielded from ionising radiation. This means that in this run gas can only cool below $1000 \, \rm{K}$ and form stars at high densities ($n_{\rm{H, tot}} \ga 10 \, \rm{cm}^{-3}$). Furthermore, the high-density gas remains much warmer than in the corresponding run with self-shielding (`UVB'), so the Jeans scale in this gas is larger. Hence the gas morphology is smoother when we do not include self-shielding. 

The simulations in Figs.~\ref{gasDensityFig} and \ref{gasTemperatureFig} were run with the full non-equilibrium chemical model. In the simulations evolved with cooling rates in chemical equilibrium, we find very similar morphologies to those seen in Figs.~\ref{gasDensityFig} and \ref{gasTemperatureFig}. 

The trends of star formation rate with metallicity and radiation field can be understood more clearly in plots of the star formation rate surface density, $\Sigma_{\rm{SFR}}$, versus gas surface density of neutral (atomic and molecular) hydrogen, $\Sigma_{\rm{HI + H_{2}}}$, i.e. the Kennicutt-Schmidt relation \citep{kennicutt98}. These are shown in Fig.~\ref{KS} for different metallicities (top row) and different radiation fields (bottom row). The intensity of the blue-green colour indicates the mass-weighted distribution of gas as a function of $\Sigma_{\rm{SFR}}$ and $\Sigma_{\rm{HI + H_{2}}}$ in simulations evolved with the full non-equilibrium chemical model. We measure these on a two-dimensional grid of cells, each $100 \, \rm{pc}$ across, that span a square region $8 \, \rm{kpc}$ across centred on the disc and are aligned such that the disc is viewed face on. We include all particles within $\pm 2 \, \rm{kpc}$ of the mid-plane of the disc in the vertical direction. We calculate $\Sigma_{\rm{SFR}}$ from the gas particles in the cell. For each gas particle that meets our star formation criteria ($n_{\rm{H_{tot}}} > 1 \, \rm{cm}^{-3}$ and $T < 1000 \, \rm{K}$), we compute the star formation rate using equation~\ref{sfr_eqn} for the star formation rate density, multiplied by the volume of the particle ($m_{\rm{gas}} / \rho_{\rm{gas}}$). We then sum the star formation rates of all star-forming particles in the cell to obtain the total instantaneous star formation rate, which we divide by the area of the cell to get $\Sigma_{\rm{SFR}}$. 

In each simulation we have combined measurements of $\Sigma_{\rm{SFR}}$ and $\Sigma_{\rm{HI + H_{2}}}$ from all snapshot outputs from $100 \, \rm{Myr}$ to $1000 \, \rm{Myr}$, at intervals of $100 \, \rm{Myr}$. The cyan curves in Fig.~\ref{KS} show the median value of $\Sigma_{\rm{SFR}}$ in bins of $\Sigma_{\rm{HI + H_{2}}}$ for simulations evolved with the full non-equilibrium model (`NonEq'; solid curves) and with cooling rates in chemical equilibrium (`Eqm'; dashed curves), and the contours show the observed Kennicutt-Schmidt relations in various samples of galaxies, as indicated in the legend. 

At high gas surface densities, $\Sigma_{\rm{SFR}}$ tends towards a power-law relation, with a slope similar to (albeit slightly larger than) that measured by \citet{kennicutt98} (the red region in Fig.~\ref{KS}). At low gas surface densities, the star formation rate drops below this power-law relation, and is cut off below a threshold gas surface density, $\Sigma_{\rm{HI + H_{2}}}^{\rm{thresh}}$. 

In the top right panel of Fig.~\ref{KS}, we see that the simulation at solar metallicity agrees fairly well with the relation measured for nearby galaxies by \citet{bigiel08} (black and grey contours). Note that we calibrated the parameters of our star formation and stellar feedback models to match the normalisation of the observed relation. The threshold $\Sigma_{\rm{HI + H_{2}}}^{\rm{thresh}}$ below which star formation is cut off increases with decreasing metallicity, as predicted by \citet{schaye04}. If we define $\Sigma_{\rm{HI + H_{2}}}^{\rm{thresh}}$ as the surface density at which the median $\Sigma_{\rm{SFR}}$ is 1 dex below the power-law relation of \citet{kennicutt98}, we find $\Sigma_{\rm{HI + H_{2}}}^{\rm{thresh}} \propto Z^{-0.3}$. This agrees well with the metallicity dependence derived by \citet{schaye04} for the column density at which the cold ISM phase forms (see his equation 23). 

In the lowest-metallicity simulation, in the top left panel of Fig.~\ref{KS}, the gas surface densities do not extend much above $\Sigma_{\rm{HI + H_{2}}}^{\rm{thresh}}$, so we cannot see whether the relation turns over and follows a power-law relation at higher gas surface densities. 

Comparing runs at ten per cent solar metallicity for different radiation fields (top centre, bottom left and bottom centre panels of Fig.~\ref{KS}), we see that, as we decrease the strength of the UV radiation field, the threshold gas surface density, $\Sigma_{\rm{HI + H_{2}}}^{\rm{thresh}}$, below which star formation is cut off decreases, as predicted by \citet{schaye04}. This is similar to the effect of increasing the metallicity that we see in the top row. We find $\Sigma_{\rm{HI + H_{2}}}^{\rm{thresh}} \propto G_{0}^{0.3}$, in good agreement with the dependence on UV intensity derived by \citet{schaye04} for the column density at which the cold ISM phase forms (his equation 23). 

The bottom right panel shows that, when we do not include self-shielding, $\Sigma_{\rm{HI + H_{2}}}^{\rm{thresh}}$ increases. Furthermore, at gas surface densities above $\Sigma_{\rm{HI + H_{2}}}^{\rm{thresh}}$, the star formation rate continues to rise steeply, and lies above the observed relation of \citet{kennicutt98}. However, since we calibrated the parameters of our star formation and stellar feedback models to reproduce the observed Kennicutt-Schmidt relation in simulations that included self-shielding, this could possibly be remedied by re-calibrating these parameters. 

\begin{figure}
\centering
\mbox{
	\includegraphics[width=84mm]{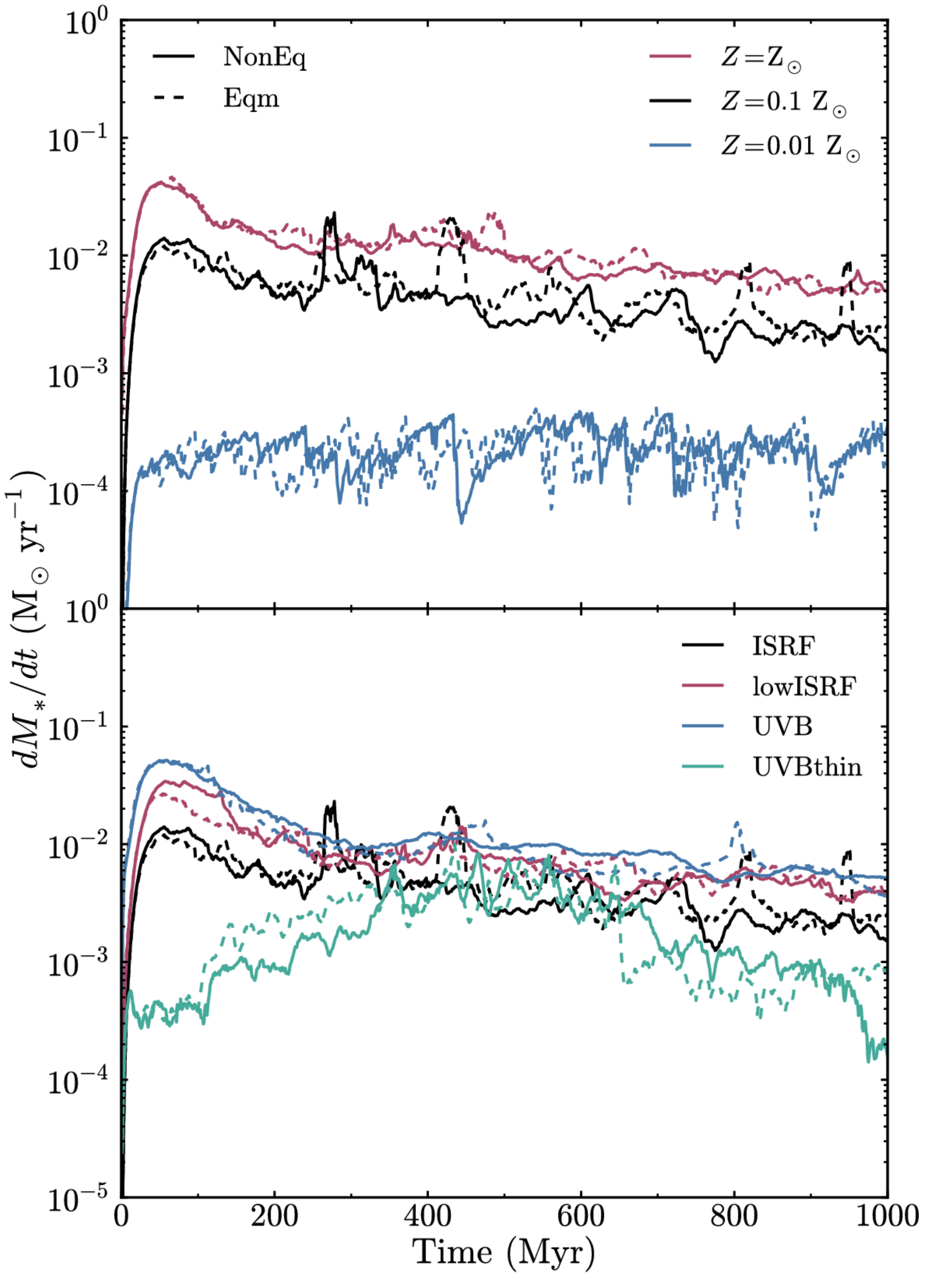}}
\caption{Star formation histories of simulations run with the full non-equilibrium chemical model \textit{(solid curves)} and run with cooling rates in chemical equilibrium \textit{(dashed curves)} at different metallicities \textit{(top panel)} and for different radiation fields \textit{(bottom panel; see Table~\ref{uv_table})}. The black curves in the two panels are from the same simulation. We find higher star formation rates in the simulations at higher metallicity and, to a lesser extent, in the presence of a weaker UV radiation field.} 
\label{SFH}
\end{figure}

The star formation histories are shown in Fig.~\ref{SFH} for different metallicities (top panel) and different radiation fields (bottom panel). The solid and dashed curves show simulations evolved with the full non-equilibrium chemical model and with cooling rates in chemical equilibrium respectively. The star formation rate initially rises rapidly as gas cools from its initial temperature of $10^{4} \, \rm{K}$ to form a cold, star-forming ISM phase. Once stellar feedback takes effect (after $\sim 50 \, \rm{Myr}$), the star formation rate levels off and then steadily declines over the course of the simulation, as gas is either consumed by star formation or driven out of the disc in outflows. 

The star formation rate increases with metallicity and is about two orders of magnitude higher for solar metallicity than for $0.01 \, \rm{Z}_{\odot}$. As we decrease the strength of the radiation field, the total star formation rate increases, by a factor of $\sim 3$ for the radiation fields that we consider here. This is similar to the trend that we see when we increase the metallicity from $0.01 \, \rm{Z}_{\odot}$ to $\rm{Z}_{\odot}$, although the size of the effect is smaller than when we vary the metallicity over this range. 

In the run without self-shielding, the total star formation rate is typically lower, by up to an order of magnitude, than in the corresponding run with self-shielding, although they are similar (to within a factor of two) between $400$ and $600 \, \rm{Myr}$. 

Comparing the solid and dashed curves in Figs.~\ref{KS} and \ref{SFH}, we see that non-equilibrium cooling has no noticeable systematic effect on the simulated Kennicutt-Schmidt relation or on the total star formation rate of the simulated galaxy. However, this conclusion may be dependent on the resolution of our simulations. In particular, our star formation prescription allows gas to form stars at densities $n_{\rm{H}} > 1.0 \, \rm{cm}^{-3}$ and temperatures $T < 1000 \, \rm{K}$. In other words, gas in our models becomes star forming once it transitions from the Warm Neutral Medium (WNM) to the Cold Neutral Medium (CNM). It is possible that there are still important non-equilibrium effects on smaller scales than we resolve that could affect the star formation rate. 

\subsection{Outflows}\label{outflows_section}

\begin{figure*}
\centering
\mbox{
	\includegraphics[width=56mm]{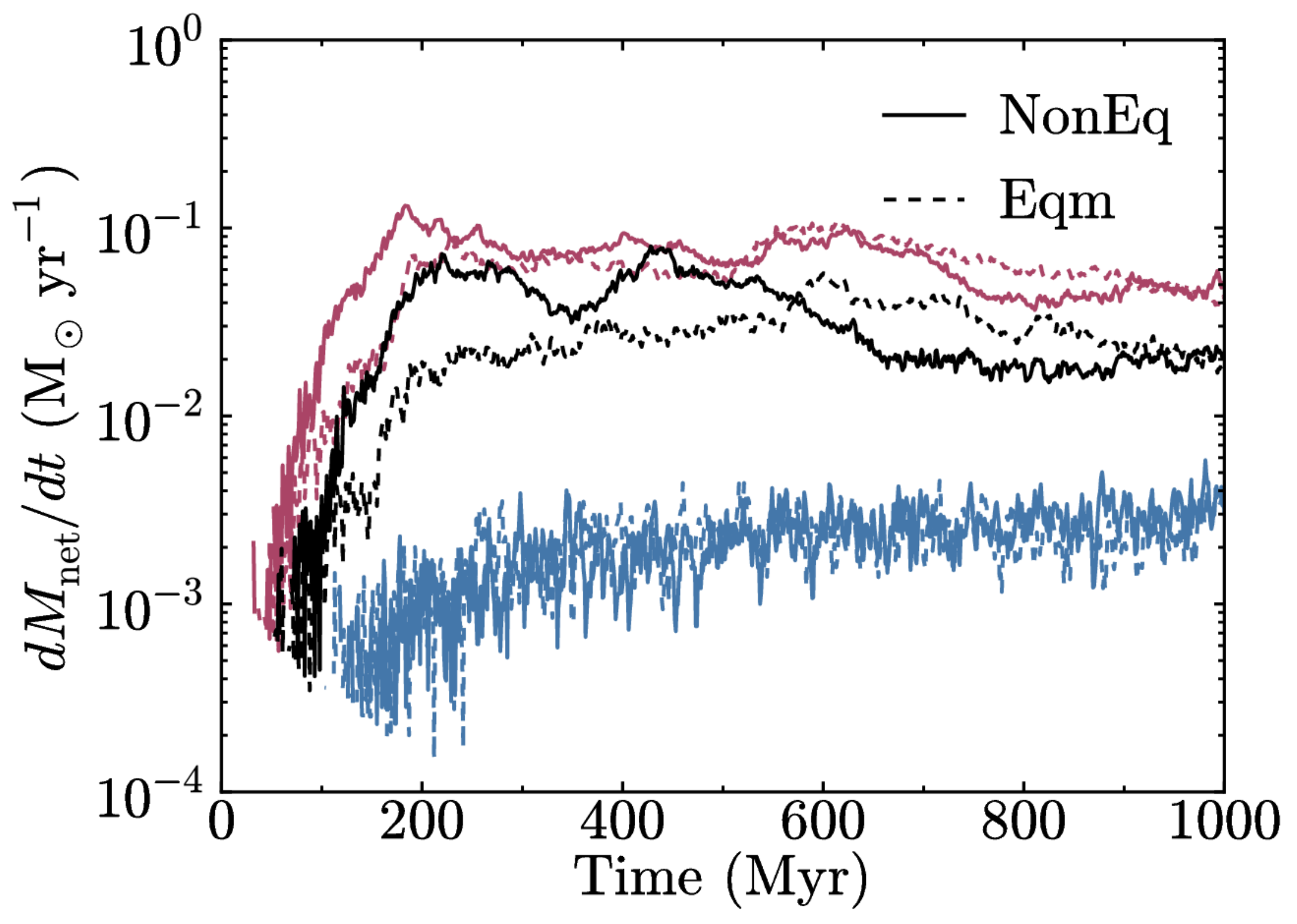}
	\includegraphics[width=56mm]{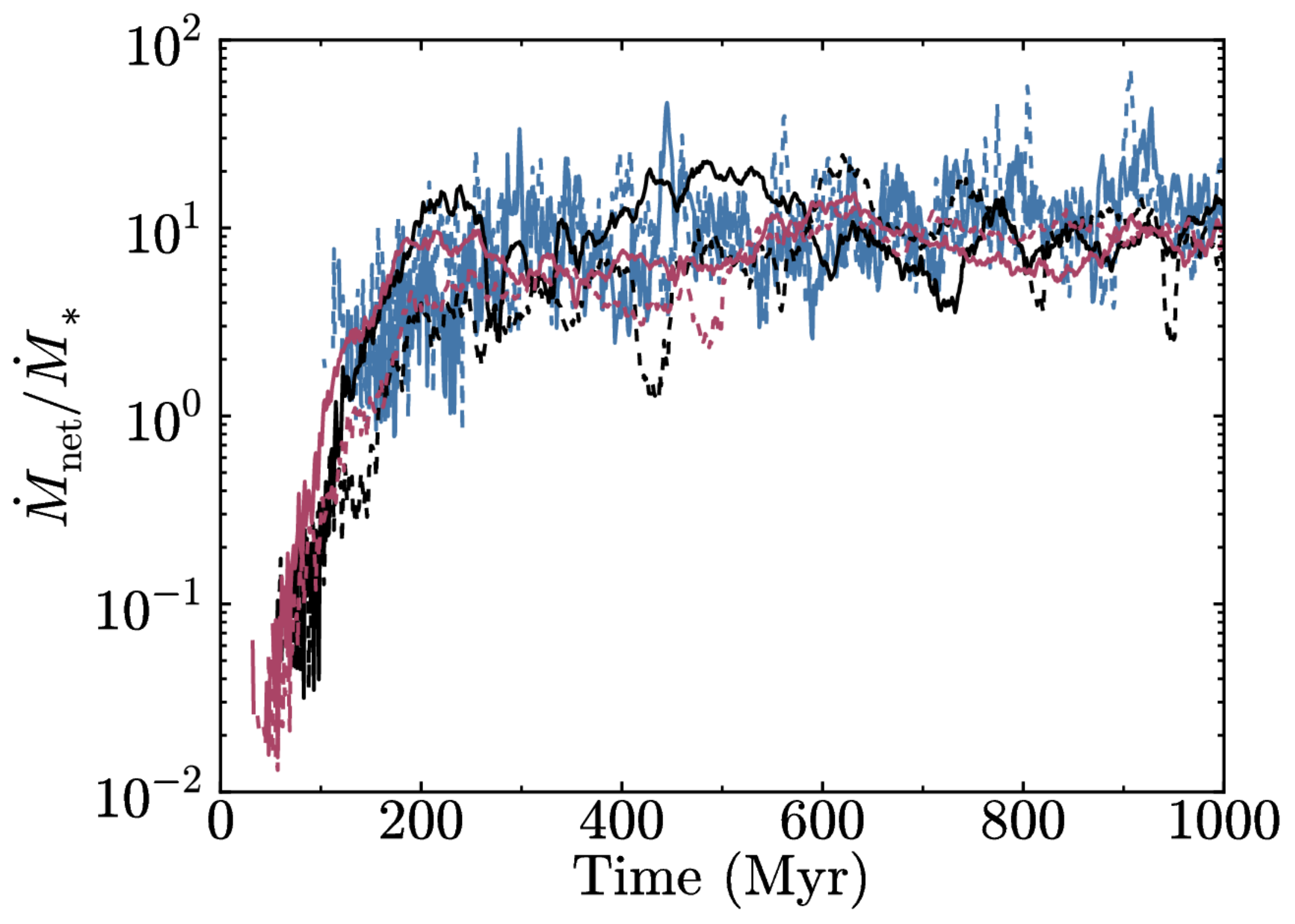}
	\includegraphics[width=56mm]{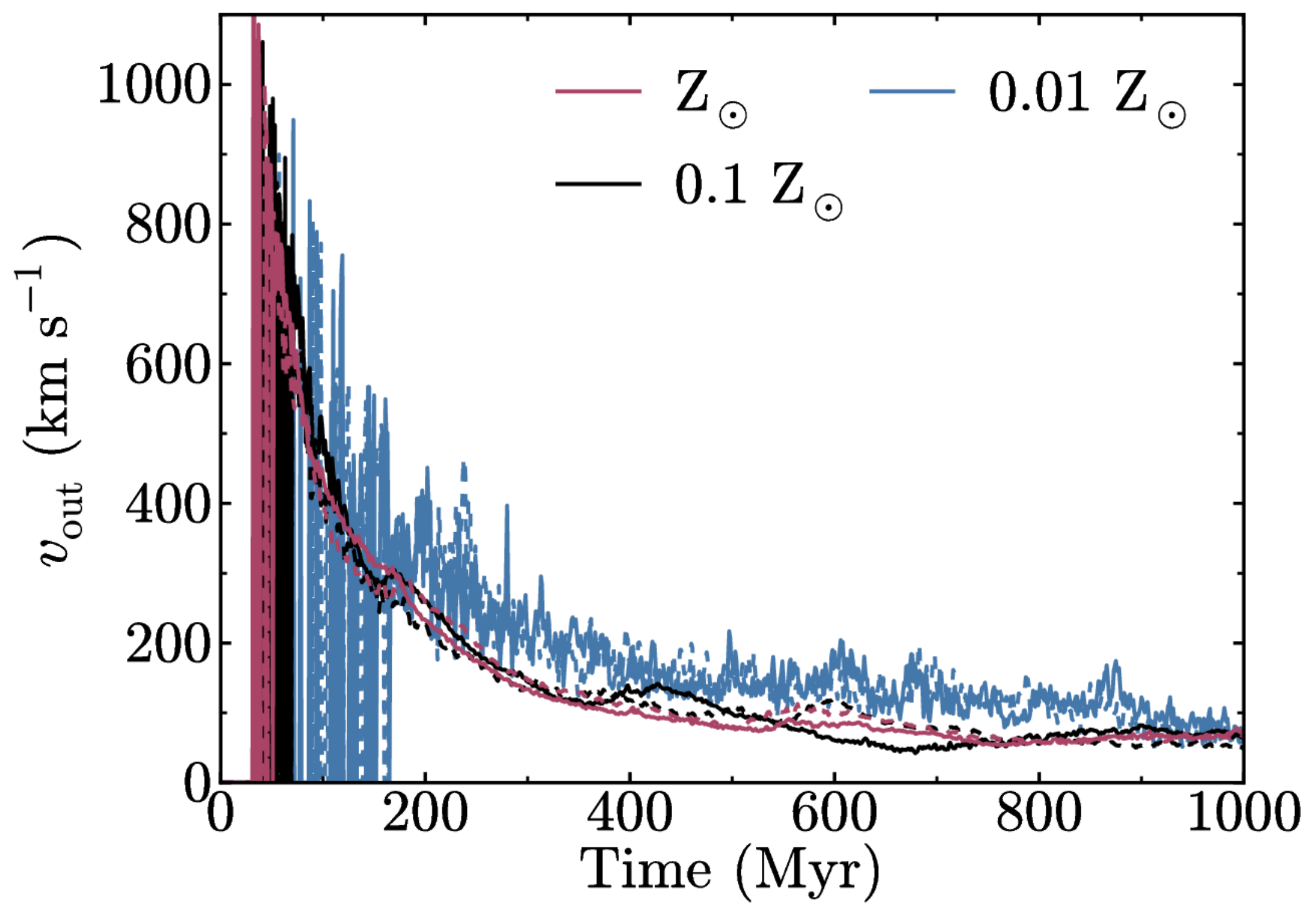}}
\mbox{
	\includegraphics[width=56mm]{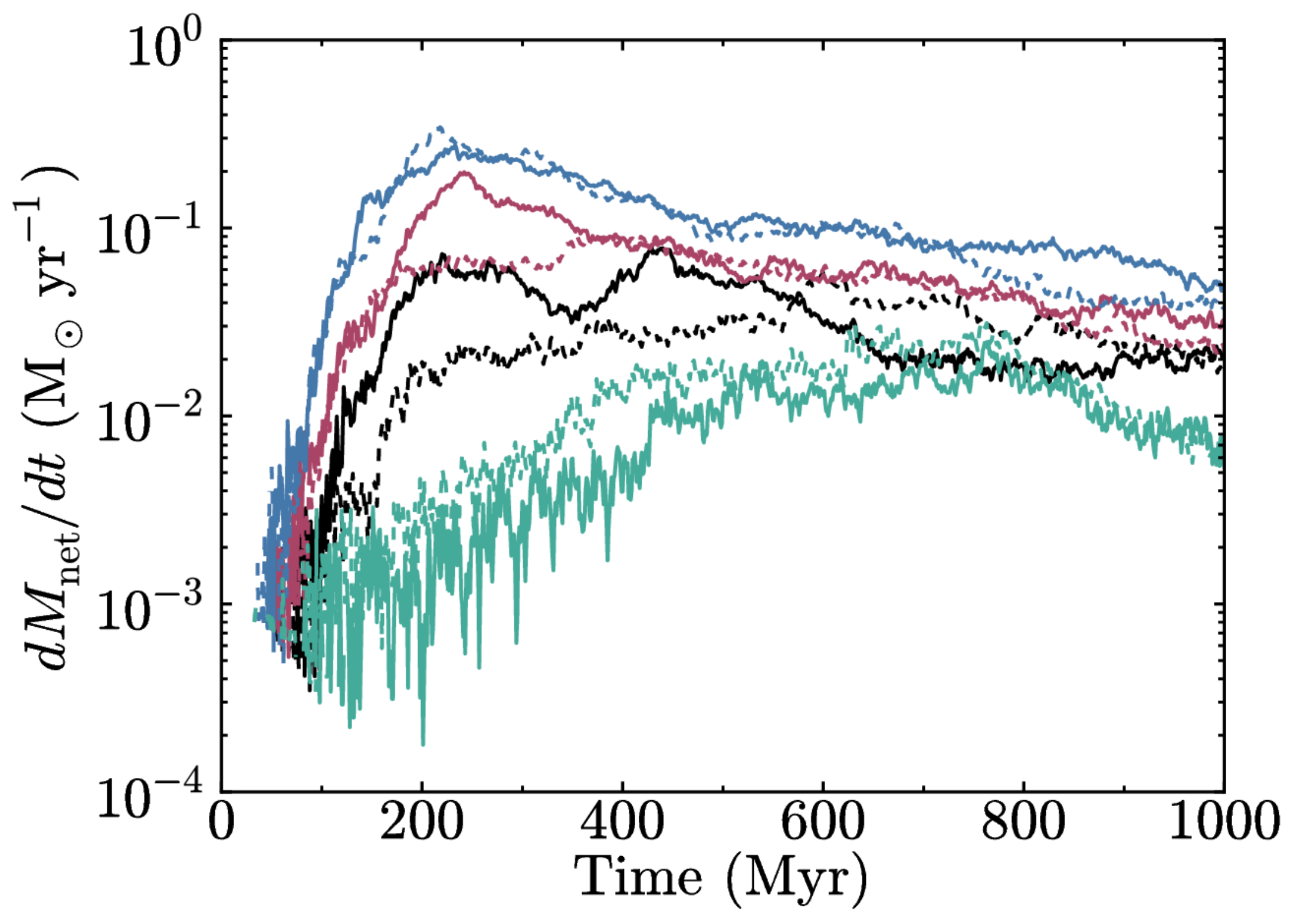}
	\includegraphics[width=56mm]{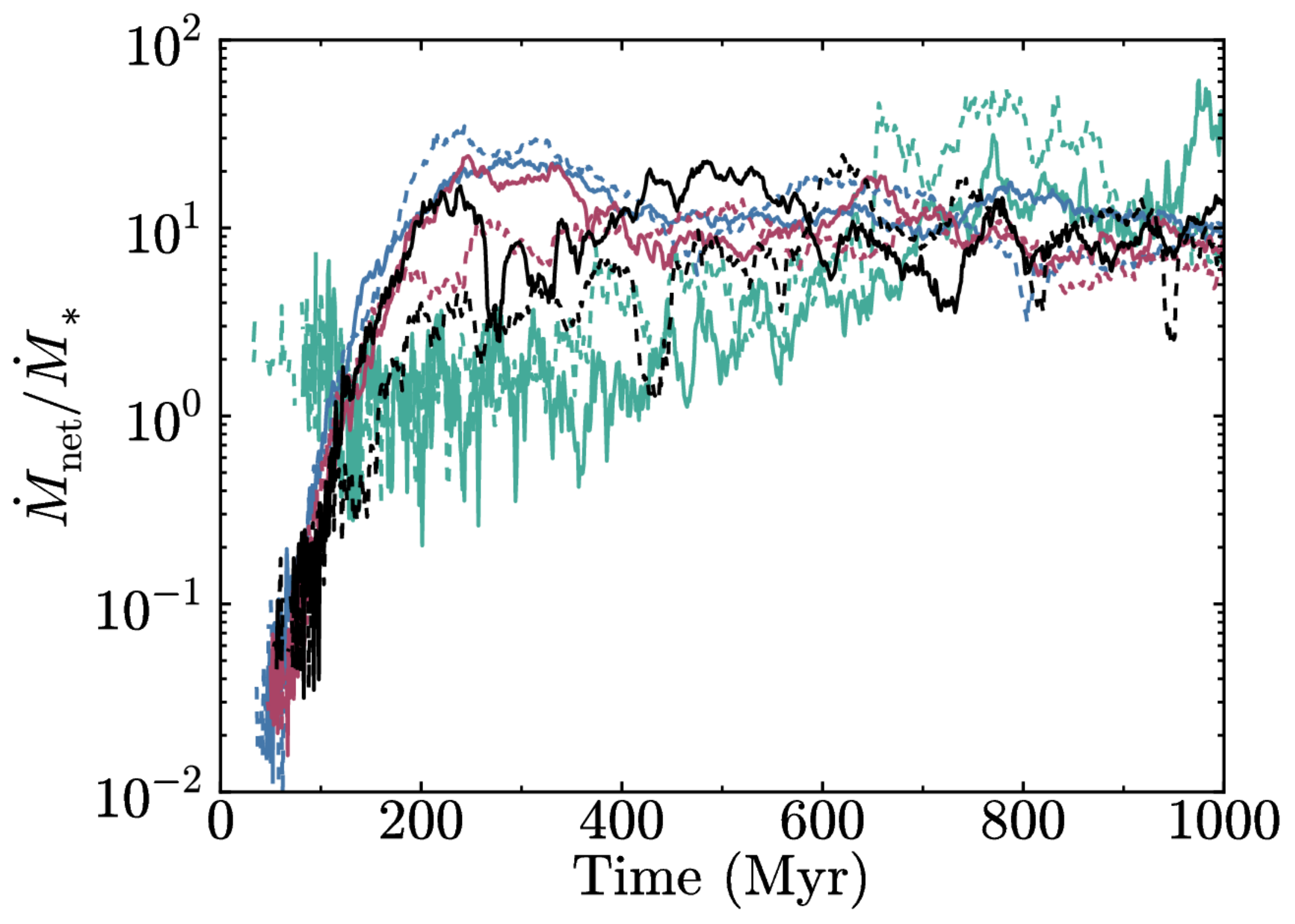}
	\includegraphics[width=56mm]{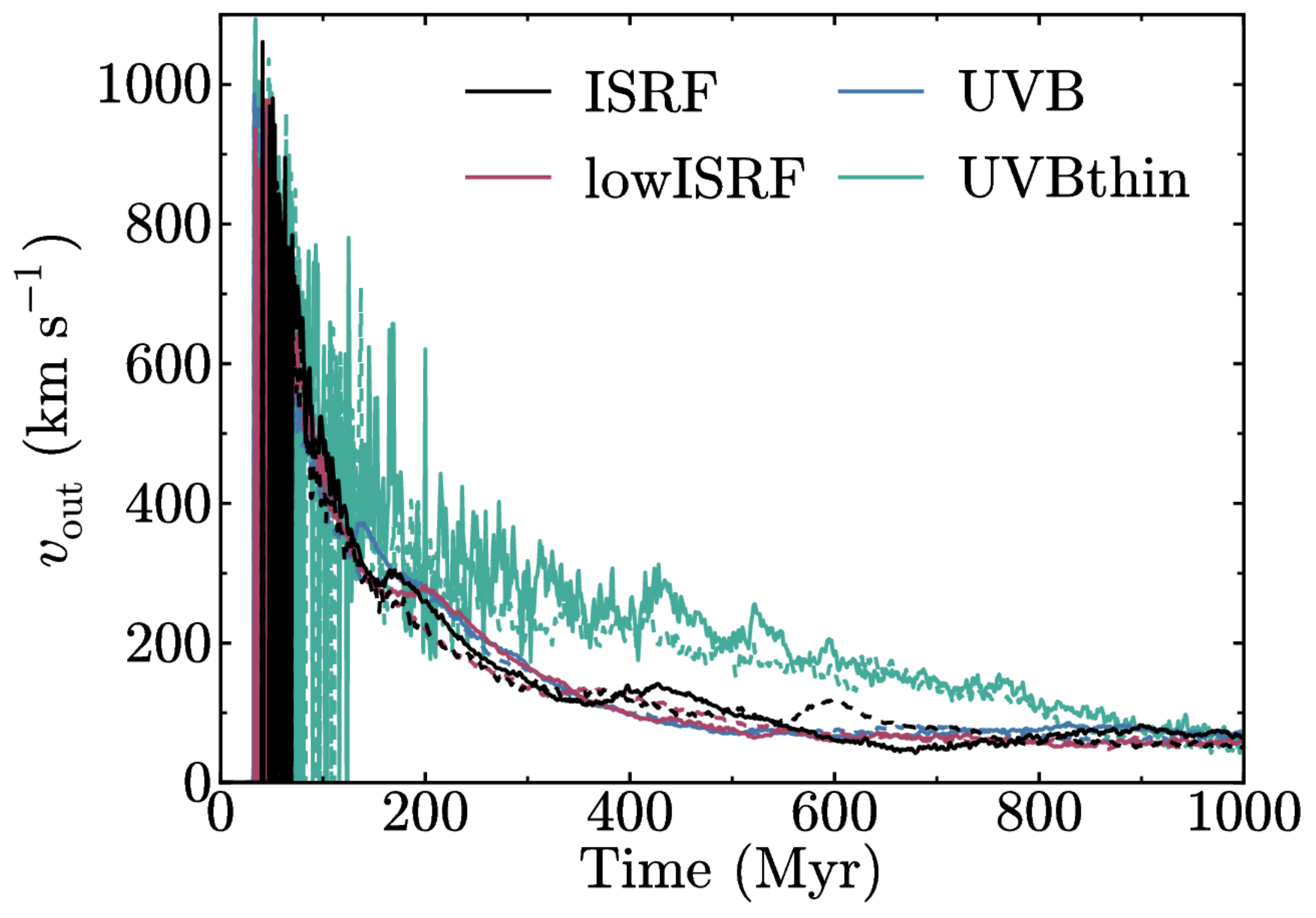}}
\caption{Evolution of the net mass outflow rate \textit{(left panels)}, mass loading factor \textit{(centre panels)} and mass-weighted mean radial outflow velocity \textit{(right panels)} measured through a thin spherical shell of radius $0.2 R_{200, \rm{crit}} = 19 \, \rm{kpc}$ and thickness $0.64 \, \rm{kpc}$. We show simulations evolved using the full non-equilibrium chemical model \textit{(solid curves)} or using cooling rates in chemical equilibrium \textit{(dashed curves)} at different metallicities \textit{(top row)} and for different radiation fields \textit{(bottom row; see Table~\ref{uv_table})}. The mass outflow rate increases with increasing metallicity and decreasing radiation field, although the mass loading factor at late times is $\sim 10$, regardless of metallicity or radiation field.}
\label{radial_outflows_fig}
\end{figure*}

Stellar feedback drives gas out of the disc in our simulations. To measure the radial mass outflow rates and velocities, we consider a spherical shell of radius $0.2 R_{200, \rm{crit}} = 19 \, \rm{kpc}$ and thickness $\Delta r = R_{200, \rm{crit}} / 150 = 0.64 \, \rm{kpc}$, centred on the origin (which is defined as the centre of the static dark matter potential). The net mass outflow rate, $dM_{\rm{net}} / dt$, is then given by: 

\begin{equation}\label{radial_mass_outflow}
\frac{dM_{\rm{net}}}{dt} = \frac{1}{\Delta r} \sum_{i = 1}^{N^{\rm{shell}}} m_{i} \frac{\mathbf{v}_{i} \cdot \mathbf{r}_{i}}{\lvert \mathbf{r}_{i} \rvert}, 
\end{equation} 
where $m_{i}$, $\mathbf{v}_{i}$ and $\mathbf{r}_{i}$ are the mass, velocity and position of the $i^{th}$ particle respectively, and we sum over the $N^{\rm{shell}}$ gas particles in the shell. We also calculate the mass-weighted mean radial velocity, $v_{\rm{out}}$, of the outflowing particles as: 

\begin{equation}\label{radial_outflow_velocity}
v_{\rm{out}} = \left< v \right>_{+} = \frac{\sum_{i = 1}^{N_{+}^{\rm{shell}}} m_{i} \left( \mathbf{v}_{i} \cdot \mathbf{r}_{i} / \lvert \mathbf{r}_{i} \rvert \right)_{+}}{\sum_{i = 1}^{N_{+}^{\rm{shell}}} m_{i}}, 
\end{equation}
where the subscript `$+$' indicates that we only consider particles with $\mathbf{v}_{i} \cdot \mathbf{r}_{i} > 0$, i.e. that are moving radially outward. 

We could split the net outflow rate in equation~\ref{radial_mass_outflow} into separate outflow and inflow rates, using particles that are moving radially outward or inward respectively. However, we find that the inflow rates at this radius are negligible in our simulations. 

We measure the mass outflow rates and mean outflow velocities from each simulation in snapshots output at $1 \, \rm{Myr}$ intervals, using equations~\ref{radial_mass_outflow} and \ref{radial_outflow_velocity}. These are shown in Fig.~\ref{radial_outflows_fig} for the simulations with different metallicities (top row), and in the presence of different UV radiation fields (bottom row). Solid and dashed curves are from simulations run with the full non-equilibrium chemical model and cooling rates in chemical equilibrium respectively. 

We see that the mass outflow rates (left panels) generally increase during the first $200 \, \rm{Myr}$, as there is a delay between the onset of star formation and the first supernovae, and it takes a finite time for the gas to reach the radius of $19 \, \rm{kpc}$ where we measure the outflows ($\sim 10 \, \rm{Myr}$ for gas travelling at $200 \, \rm{km} \, \rm{s}^{-1}$). In the simulations without self-shielding (green curves in the bottom row), this initial increase is more gentle and extends over a longer period of time ($\sim 500 \, \rm{Myr}$). We saw in Fig.~\ref{SFH} that the star formation rates in the simulations without self-shielding also increase more gently over this period. After the initial rise, the mass outflow rates fluctuate around an approximately steady or gently declining value. 

The outflow rates tend to be larger in simulations at higher metallicity or in the presence of a weaker UV radiation field, due to the larger star formation rates that we find in these cases (see Fig.~\ref{SFH}). In the centre panels of Fig.~\ref{radial_outflows_fig} we show the evolution of the ratio between the mass outflow rate and the star formation rate, i.e. the mass loading factor. After $200 \, \rm{Myr}$, the mass loading factor tends towards a value of $\sim 10$, independent of the metallicity or strength of the UV radiation field. An exception to this trend is seen in the simulations that do not include self-shielding (green curves in the bottom row), which show a steady rise in the mass loading factor from $\sim 1$ at $200 \, \rm{Myr}$ to $\sim 30$ at the end of the simulation, albeit with large fluctuations on timescales of a few $\rm{Myr}$. 

One caveat is that we do not include a gaseous halo in the initial conditions. In a realistic galaxy, the outflowing gas may be slowed down as it interacts with an existing gaseous halo, which could decrease the mass loading factor at $R_{200, \rm{crit}}$. Alternatively, it may sweep up more material from the halo, which could increase the mass loading factor. 

In the right panels of Fig.~\ref{radial_outflows_fig}, we show the evolution of the mean radial velocity of outflowing gas. Initially, we find very large outflow velocities ($v_{\rm{out}} \sim 1000 \, \rm{km} \, \rm{s}^{-1}$), because the fastest particles ejected from the disc are the first to reach the radius at which we measure the outflows. The mean outflow velocity subsequently decreases as the slower particles reach this radius, and tends towards a value of $\sim 65 \, \rm{km} \, \rm{s}^{-1}$ by the end of the simulation. 

Comparing solid and dashed curves in Fig.\ref{radial_outflows_fig}, we see that the outflows are generally not strongly affected by the use of equilibrium cooling rates. In the runs at solar and ten per cent solar metallicity in the presence of the \citet{black87} ISRF (red and black curves respectively in the top row of Fig.~\ref{radial_outflows_fig}), the mass outflow rates (left panels) rise more gently in the first $200 \, \rm{Myr}$ in the simulations run with equilibrium cooling rates (dashed curves), compared to the corresponding runs evolved with non-equilibrium cooling rates (solid curves). However, this difference becomes less pronounced in the presence of a weaker UV radiation field (red and blue curves in the bottom row of Fig.~\ref{radial_outflows_fig}). Also, the mass loading factors (centre panels) and outflow velocities (right panels) are similar whether we use non-equilibrium or equilibrium cooling. 

We also considered how the mass loading factor, $\beta = \dot{M}_{\rm{out}} / \dot{M}_{\ast}$, depends on gas surface density in our simulations. To do this, we measured vertical outflows close to the disc, at a vertical height of $1 \, \rm{kpc}$ above and below the mid-plane of the disc. We positioned two thin sheets of thickness $\Delta z = 50 \, \rm{pc}$ parallel to the disc (in the $x-y$ plane) at a vertical distance $\lvert z \rvert = 1 \, \rm{kpc}$. Each sheet covered a square region $8 \, \rm{kpc}$ across centred on the galaxy centre, which corresponds to the region shown in the face-on views in Figs.~\ref{gasDensityFig} and \ref{gasTemperatureFig}. The vertical mass outflow rate through these two sheets is given by:

\begin{equation}\label{vertical_outflow_rate}
\frac{d M_{\rm{out}}}{dt} = \frac{1}{\Delta z} \sum_{i = 1}^{N_{+}^{\rm{sheet}}} m_{i} \lvert v_{z, i} \rvert_{+}, 
\end{equation}
where $v_{z, i}$ is the $z$-component of the velocity of the $i^{\rm{th}}$ particle, the subscript `$+$' indicates that we only include particles with $\mathbf{z} \cdot \mathbf{v} > 0$ (i.e. that are moving vertically away from the mid-plane), and we sum over the $N_{+}^{\rm{sheet}}$ gas particles with $\mathbf{z} \cdot \mathbf{v} > 0$ in the two sheets. 

To compare the mass loading factor at a vertical height of $1 \, \rm{kpc}$ to the gas surface density within the disc, we divided the two thin sheets into a grid of cells, each $100 \, \rm{pc}$ across. We then measured the mass outflow rates through the two sheets in each cell, along with the star formation rate and gas surface density in the disc within the cell. 

\begin{figure}
\centering
\mbox{
	\includegraphics[width=84mm]{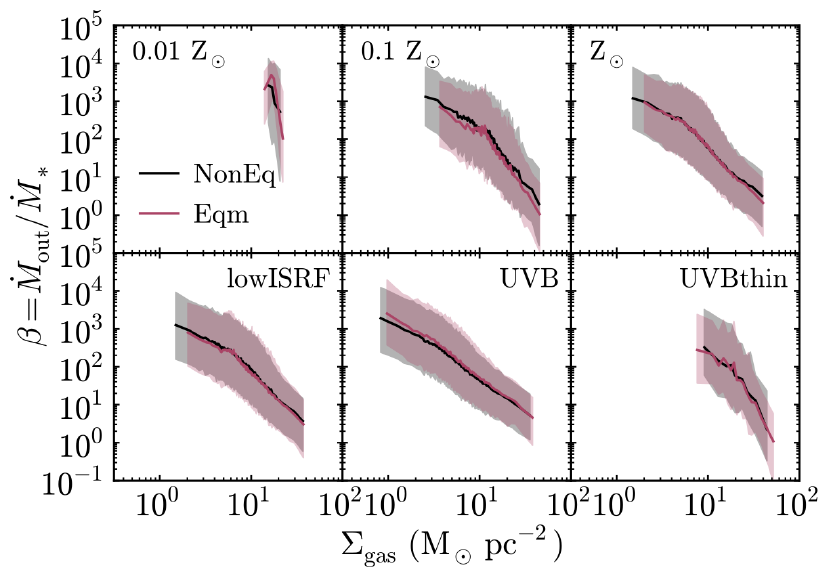}}
\caption{Mass loading factor, $\beta = \dot{M}_{\rm{out}} / \dot{M}_{\ast}$, plotted against gas surface density, $\Sigma_{\rm{gas}}$, measured in regions of the disc $100 \, \rm{pc}$ across at a vertical distance $\lvert z \rvert$ of $1 \, \rm{kpc}$ above and below the mid-plane of the disc. We show simulations at different metallicities \textit{(top row)} and for different radiation fields \textit{(bottom row; see Table~\ref{uv_table})}, evolved using the full non-equilibrium chemical model \textit{(black)} or using cooling rates in chemical equilibrium \textit{(red)}. The solid curves show the median mass loading factor in bins of $\Sigma_{\rm{gas}}$, and the shaded regions indicate the range between the tenth and ninetieth percentiles in each bin. The mass loading factor decreases with increasing $\Sigma_{\rm{gas}}$, and this relation is unaffected by using equilibrium cooling.}
\label{mass_loading_fig}
\end{figure}

In Fig.~\ref{mass_loading_fig} we plot the mass loading factor as a function of gas surface density for different metallicities (top row) and different UV radiation fields (bottom row). We measured the mass loading factors and gas surface densities in all snapshot outputs from each simulation, at intervals of $1 \, \rm{Myr}$, and we show the median mass loading factor (solid curves) and the range between the tenth and ninetieth percentiles (shaded regions) in bins of gas surface density. Simulations run with the full non-equilibrium and equilibrium chemical models are shown in black and red respectively. 

The mass loading factor decreases with increasing gas surface density, following approximately a power-law, with $\beta \propto \Sigma_{\rm{gas}}^{-2}$. At constant gas surface density, the mass loading factor increases with decreasing metallicity and increasing UV radiation field. The mass loading factors in the different panels of Fig.~\ref{mass_loading_fig} suggest a scaling of $\beta \propto Z^{-0.5}$ and $\beta \propto G_{0}^{0.3}$, although these scalings are highly uncertain, as we only consider a small number of different metallicities and radiation fields. 

Comparing black and red curves in Fig.~\ref{mass_loading_fig}, we see that we recover the same relation between mass loading factor and gas surface density, regardless of whether we use non-equilibrium cooling rates or cooling rates in chemical equilibrium. 

\citet{hopkins12} explored the dependence of $\beta$ on galaxy properties in their SPH simulations of isolated galaxies. They found $\beta = 10 (V_{\rm{C}}(R) / 100 \, \rm{km} \, \rm{s}^{-1})^{-1} (\Sigma_{\rm{gas}}(R) / 10 \, \rm{M}_{\odot} \, \rm{pc}^{-2})^{-0.5}$ (their equation 8), where $V_{\rm{C}}$ is the circular velocity at radius $R$. This has a much weaker dependence on $\Sigma_{\rm{gas}}$, with a power-law slope of $-0.5$, compared to $\sim -2$ from our simulations. This different scaling with $\Sigma_{\rm{gas}}$ may be due to the different prescription for stellar feedback that was used by \citet{hopkins12}. 

The relation betweem $\beta$ and $\Sigma_{\rm{gas}}$ has also been explored by \citet{creasey13}. They used high-resolution mesh simulations of supernova-driven galactic winds in a $1 \, \rm{kpc}$ column through a galactic disc, but they did not include radiative cooling below $10^{4} \, \rm{K}$. They showed that $\beta$ increases with decreasing $\Sigma_{\rm{gas}}$, and with increasing gas fraction, $f_{\rm{gas}}$. They fit a power-law dependence of $\beta$ on $\Sigma_{\rm{gas}}$ and $f_{\rm{gas}}$ in their simulations, and they found $\beta = 13 \Sigma_{\rm{gas}}^{-1.15} f_{\rm{gas}}^{0.16}$ (see equations 39-42 of \citealt{creasey13}). This power-law scaling with $\Sigma_{\rm{gas}}$ is intermediate between our value and the relation of \citet{hopkins12}. 

We find much larger mass loading factors than \citet{creasey13} at a given gas surface density. For example, at $\Sigma_{\rm{gas}} = 10 \, \rm{M}_{\odot} \, \rm{pc}^{-2}$, we typically find a median mass loading factor of $\sim 100$ at $\lvert z \rvert = 1 \, \rm{kpc}$, whereas, for a gas fraction $f_{\rm{gas}} = 0.3$, the best-fit relation of \citet{creasey13} gives a mass loading factor of 0.8. For comparison, the relation of \citet{hopkins12} gives $\beta = 20$, where $V_{\rm{C}} \sim 50 \, \rm{km} \, \rm{s}^{-1}$ in our simulations. \citet{creasey13} measured the outflows from their simulation volume closer to the mid-plane, at $\lvert z \rvert = 500 \, \rm{pc}$. However, when we consider outflows at $\lvert z \rvert = 500 \, \rm{pc}$ in our simulations, we still find a mass loading factor $\sim 100$ at $\Sigma_{\rm{gas}} = 10 \, \rm{M}_{\odot} \, \rm{pc}^{-2}$, and a slope $\sim -2$.

There are several differences between our simulations and those of \citet{creasey13} that could explain these discrepancies. Most importantly, \citet{creasey13} did not include radiative cooling below $10^{4} \, \rm{K}$, whereas gas in our simulations can cool to $10 \, \rm{K}$. The presence of a cold phase in the ISM is likely to affect how outflowing gas escapes the disc. \citet{creasey13} also did not include rotation of the disc, or self-gravity of the gas. However, they did use a slightly higher resolution ($1.6 \, \rm{pc}$) than we have in our simulations (we use a gravitational softening length of $3.1 \, \rm{pc}$ for gas particles). \citet{creasey13} also injected energy from individual supernovae, whereas our stellar feedback model requires that we inject energy from several supernovae simultaneously in a single feedback event, to prevent artificial radiative losses (see section~\ref{feedback_section}). 

\subsubsection{Chemistry of outflowing gas} 

We have seen that the mass outflow rates and velocities are generally not strongly affected by non-equilibrium cooling. However, we might expect the abundances in outflowing gas to be out of equilibrium, since this gas is highly dynamic. This would be important for comparing with observations of particular chemical species in outflows. For example, molecular outflows have been observed in extragalactic systems, including starbursting galaxies and Active Galactic Nuclei (AGN). These have been observed in emission and absorption from a number of molecules, including CO, H$_{2}$, OH and HCO$^{+}$ \citep[e.g.][]{baan89, walter02, leon07, sakamoto09, sturm11, emonts14, geach14}. 

\begin{figure}
\centering
\mbox{
	\includegraphics[width=84mm]{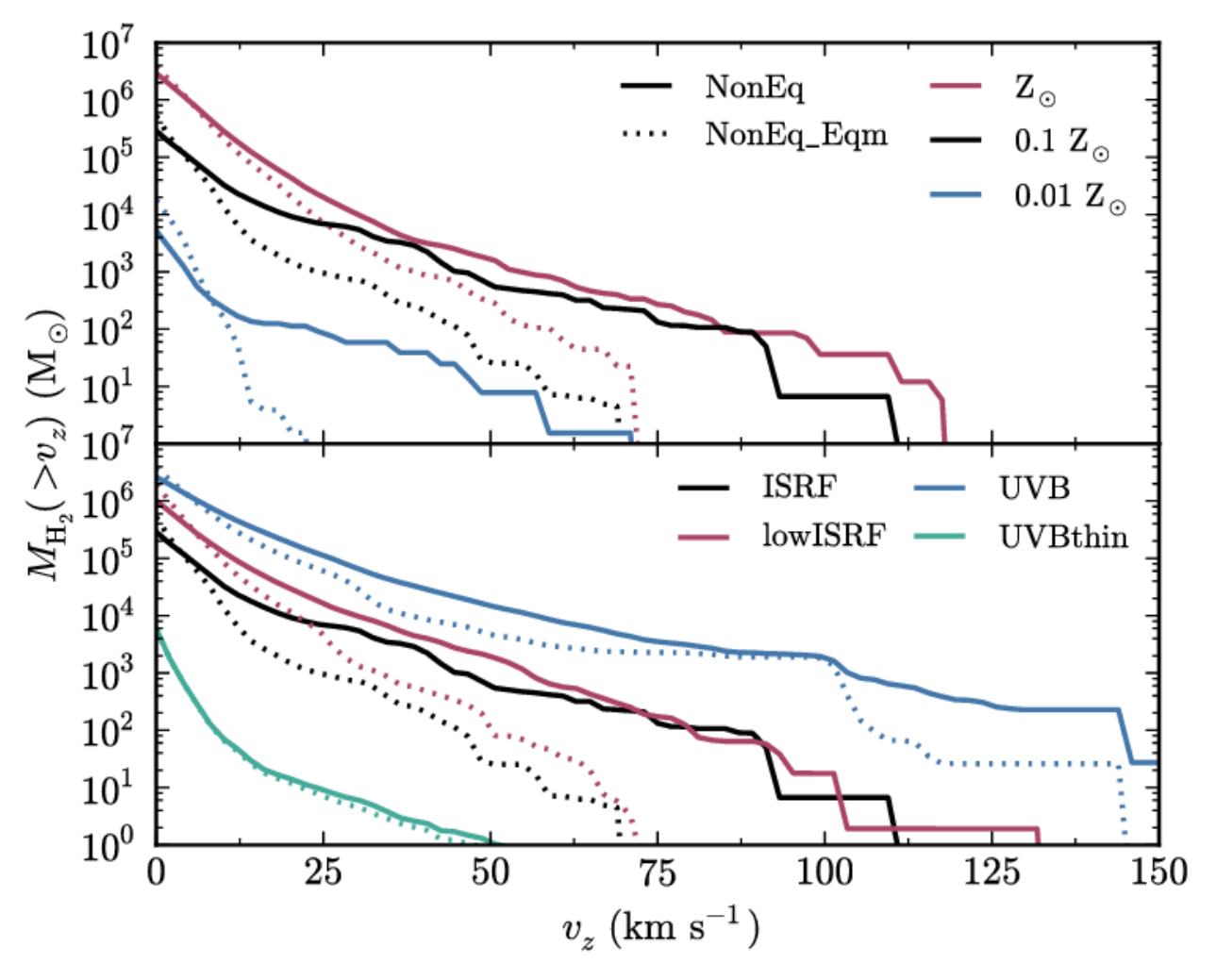}}
\caption{Mass of molecular hydrogen in particles moving away from the mid-plane of the disc with a vertical velocity $> v_{z}$, plotted as a function of $v_{z}$, for different metallicities \textit{(top panel)} and for different radiation fields \textit{(bottom panel; see Table~\ref{uv_table})}. We calculate the H$_{2}$ masses from simulations run using the full chemical model using non-equilibrium abundances \textit{(solid curves)} and from the same simulations but with abundances set to chemical equilibrium in post-processing \textit{(dotted curves)}. We average these curves over ten snapshots for each simulation, at intervals of $100 \, \rm{Myr}$. The mass of H$_{2}$ in outflowing gas ($\ga 50 \, \rm{km} \, \rm{s}^{-1}$) is generally much higher in non-equilibrium, e.g. by a factor of $\sim 20$ in the ISRF run.}
\label{H2_outflow_fig}
\end{figure}

To investigate whether the molecular abundances are out of equilibrium in outflowing gas, we measured the total mass of H$_{2}$ in particles with a vertical velocity (perpendicular to the plane of the disc) greater than some velocity $v_{z}$, i.e. $M_{\rm{H_{2}}}(> v_{z})$. We include only particles moving away from the mid-plane of the disc (i.e. that are outflowing), with $\mathbf{z} \cdot \mathbf{v} > 0$. 

In Fig.~\ref{H2_outflow_fig} we plot $M_{\rm{H_{2}}}(> v_{z})$ as a function of $v_{z}$, averaged over ten snapshot outputs for each simulation at intervals of $100 \, \rm{Myr}$. We show different metallicities and different radiation fields in the top and bottom panels respectively. The solid curves in each panel show the H$_{2}$ mass computed from the simulations evolved with the full chemical model using non-equilibrium H$_{2}$ abundances, while the dotted curves are calculated from the same non-equilibrium simulations, but using abundances that are set to equilibrium in post-processing. The latter we label as `NonEq\_Eqm', to distinguish them from the `Eqm' simulations that were evolved with cooling rates in chemical equilibrium (not shown). By comparing the `NonEq' and `NonEq\_Eqm' models in Fig.~\ref{H2_outflow_fig}, we consider the same distribution of gas in a given snapshot, and the only differences are due to the H$_{2}$ abundances being out of equilibrium. However, if we compared these with the `Eqm' simulations, we could also see differences in the outflow rate of molecular hydrogen that are caused by the different gas distribution in that simulation. For example, if there had recently been a strong burst of star formation in one snapshot that did not occur in the other simulation, we would see an enhanced molecular outflow that was not caused by non-equilibrium effects. Therefore, we do not include the `Eqm' simulations in Fig.~\ref{H2_outflow_fig}. 

At low $v_{z}$ (below $5 \, \rm{km} \, \rm{s}^{-1}$) the non-equilibrium H$_{2}$ masses are generally slightly lower than in equilibrium. The H$_{2}$ mass at low $v_{z}$ is dominated by molecular clouds in the disc and, as we will see in section~\ref{phase_structure}, H$_{2}$ is underabundant in gas that is starting to become molecular. However, at larger $v_{z}$, where we include only gas that is outflowing, we find much larger H$_{2}$ masses when we use non-equilbrium abundances than when we assume chemical equilibrium. This gas was previously in molecular clouds, but has not yet had enough time for the H$_{2}$ to be destroyed and reach the new chemical equilibrium state since being ejected. For example, in the simulation at $0.1 \, \rm{Z}_{\odot}$ in the presence of the \citet{black87} ISRF (black curves), we find $600 \, \rm{M}_{\odot}$ of molecular hydrogen outflowing at $> 50 \, \rm{km} \, \rm{s}^{-1}$, compared to only $30 \, \rm{M}_{\odot}$ if we assume chemical equilibrium. Thus, non-equilibrium chemistry enhances the mass of outflowing H$_{2}$ by a factor $\sim 20$ in this example. We see such differences in all runs with the \citet{black87} ISRF and ten per cent of the \citet{black87} ISRF, but the differences due to non-equilibrium abundances are much smaller in the presence of the \citet{haardt01} UVB (blue curves in the bottom panel). Also, in the simulation without self-shielding (green curves in the bottom panel), the non-equilibrium and equilibrium H$_{2}$ masses are almost identical, and are much lower (by more than two orders of magnitude) than in the corresponding run with self-shielding, because gas does not become shielded from the dissociating radiation in this run.

\begin{figure*}
\centering
\mbox{
	\includegraphics[width=168mm]{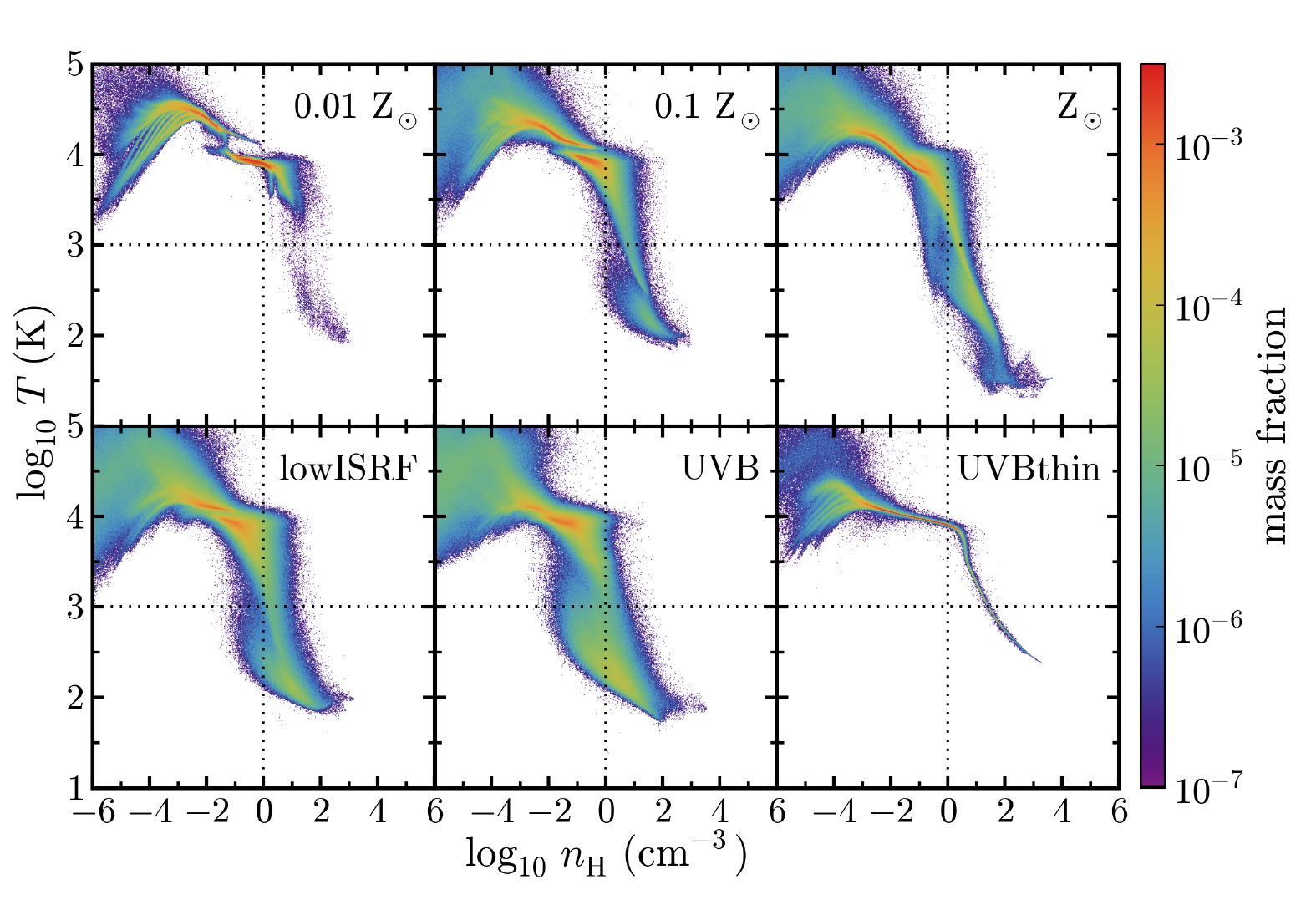}}
\caption{The distribution of gas in the temperature-density plane from simulations at different metallicities \textit{(top row)} and for different radiation fields \textit{(bottom row; see Table~\ref{uv_table})}. These simulations were run using the full non-equilibrium chemical model. In each panel we combine snapshot outputs taken at intervals of $100 \, \rm{Myr}$. The colour scale indicates the mass fraction of gas in each pixel. The dotted horizontal and vertical lines indicate the temperature and density thresholds, respectively, of our star formation prescription. Thus gas in the bottom right region of each panel is allowed to form stars. We find more cold, star forming gas in the simulations run at higher metallicity and in the presence of a weaker UV radiation field.}
\label{T_rho_fig}
\end{figure*}

Fig.~\ref{H2_outflow_fig} therefore demonstrates that non-equilibrium chemistry can be very important for modelling molecular outflows, as the molecules in gas that is ejected from the galaxy are not instantly destroyed, but instead take time to evolve to a new chemical equilibrium. 

\subsection{Phase structure of the ISM}\label{phase_structure} 

Fig.~\ref{T_rho_fig} shows two-dimensional histograms of the gas temperature and density for simulations with different metallicities (top row) and radiation fields (bottom row), evolved with the full non-equilibrium chemical model. In each panel we stack snapshot outputs taken at intervals of $100 \, \rm{Myr}$ to show the time-averaged distribution of gas. The colour scale indicates the mass fraction of gas in each pixel, so we see the mass-weighted distribution, rather than the volume-weighted distribution. 

Comparing different metallicities (top row), we see that increasing the metallicity allows more gas to cool to lower temperatures and higher densities. Reducing the strength of the radiation field (top centre, bottom left and bottom centre panels) also increases the amount of cold gas in the simulation, due to the reduced photoionisation heating and photoelectric dust heating. Note that, in the star formation prescription that we use, gas particles are allowed to form stars if they have a density $n_{\rm{H}} > 1.0 \, \rm{cm}^{-3}$ and a temperature $T < 10^{3} \, \rm{K}$, i.e. if they lie in the bottom right region delineated by the vertical and horizontal dotted lines in each panel. We thus see that there is more star forming gas in the galaxies at higher metallicity or in the presence of a weaker UV radiation field. This explains the higher star formation rates that we saw in section~\ref{sf_section}, and hence the stronger supernova-driven galactic winds that we saw in section~\ref{outflows_section}, at higher metallicity and lower radiation field strength. 

At densities $10^{-2} \, \rm{cm}^{-3} \la n_{\rm{H}} \la 10^{0} \, \rm{cm}^{-3}$, the gas at low metallicity follows two distinct tracks in the temperature-density plane, which become less distinct at higher metallicity. We find that the high-temperature track consists of strongly ionised gas, which experiences strong photoheating, while the low-temperature track consists of neutral gas that has become shielded from hydrogen-ionising radiation (see, for example, the first and third rows of Fig.~\ref{nonEqRatio_elec_fig}, which show temperature-density diagrams with a colour scale indicating the electron abundance). At high metallicities, metal cooling reduces the thermal equilibrium temperature of the ionised gas and thus brings the two tracks closer together. We also see that the distinction between these two tracks is less clear in the presence of a weaker UV radiation field, and that they coincide if we neglect self-shielding. 

In the bottom right panel of Fig.~\ref{T_rho_fig}, we see that there is much less scatter in the gas temperature and density when we do not include self-shielding of gas from the radiation field, with most of the gas at densities $n_{\rm{H}} > 10^{-1} \, \rm{cm}^{-3}$ following a very narrow region in the temperature-density plane. When we include self-shielding, gas particles at a particular density and temperature can exhibit a wide range of different shielding column densities, which explains the larger scatter that we see in the temperature-density plane when self-shielding is included. 

\begin{figure}
\centering
\mbox{
	\includegraphics[width=84mm]{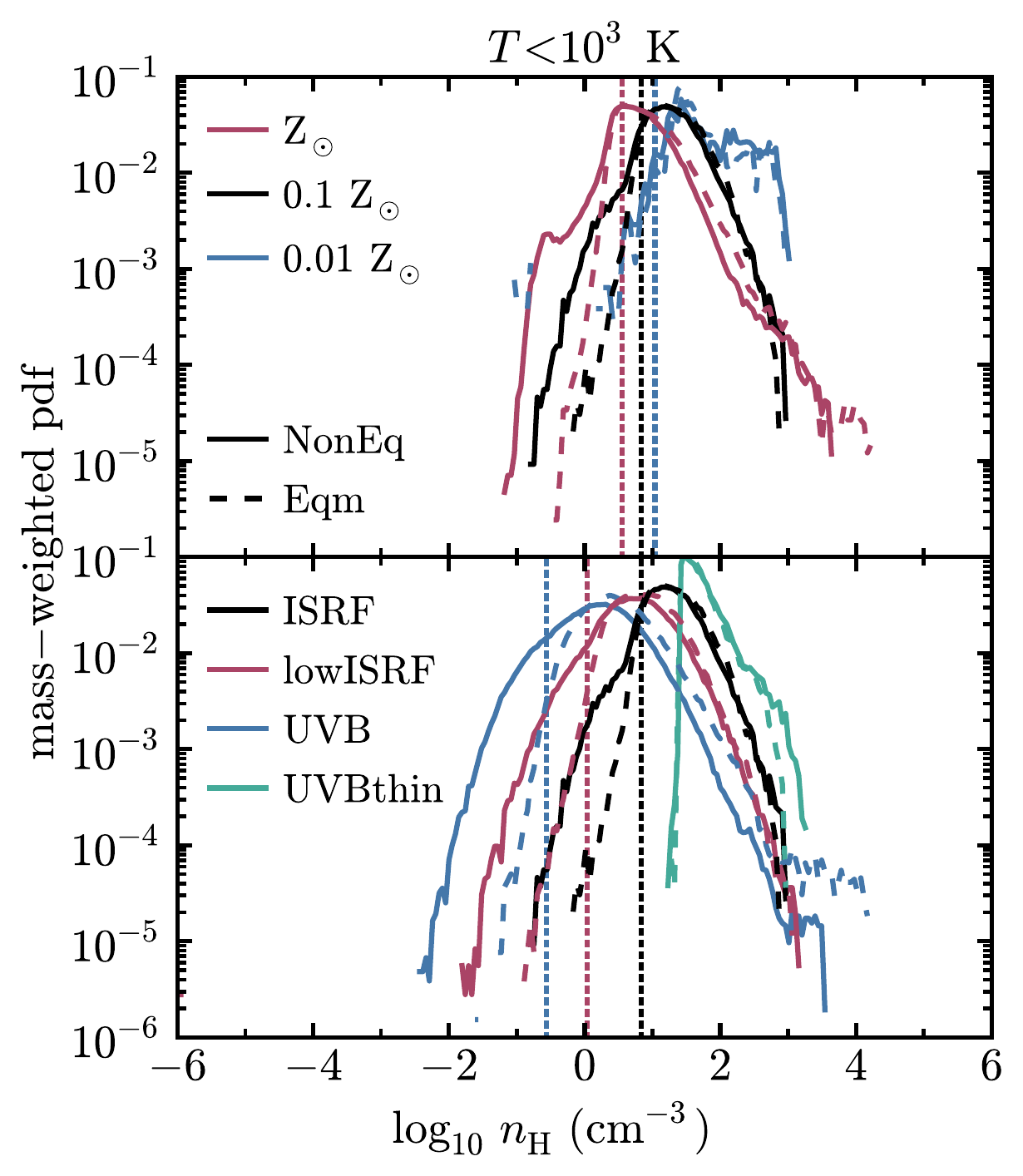}}
\caption{One-dimensional mass-weighted density pdf of cold gas ($T < 10^{3} \, \rm{K}$) from simulations evolved using the full non-equilibrium chemical model \textit{(solid curves)} and using cooling rates in chemical equilibrium \textit{(dashed curves)} at different metallicities \textit{(top panel)} and for different UV radiation fields \textit{(bottom panel; see Table~\ref{uv_table})}. The vertical dotted lines show the minimum density of the cold neutral medium predicted by the model of \citet{wolfire03}, as a function of metallicity and radiation field.}
\label{rho_pdf_fig}
\end{figure}

In the simulations evolved with equilibrium cooling rates, we generally find similar distributions to those in Fig.~\ref{T_rho_fig} for the simulations evolved with the full non-equilibrium chemical model. However, there are some regions in this plane that appear different. To quantify these differences, we show the one-dimensional probability density functions (pdfs) of gas density in Fig.~\ref{rho_pdf_fig} for simulations with different metallicities (top panel) and different radiation fields (bottom panel). Solid and dashed curves correspond to the full non-equilibrium chemical model and cooling rates in chemical equilibrium respectively. We show here the density distributions for cold gas, with $T < 10^{3} \, \rm{K}$, which corresponds to the temperature threshold below which gas particles can form stars in our star formation prescription. This temperature is indicated by the dotted horizontal lines in Fig.~\ref{T_rho_fig}. 

At metallicities $Z \geq 0.1 \, \rm{Z}_{\odot}$, we see that the density distribution in the top panel of Fig.~\ref{rho_pdf_fig} extends to lower densities, by $\sim 0.5 \, \rm{dex}$, when the galaxy is evolved using the full non-equilibrium chemical model. We also see this effect when we consider weaker UV radiation fields at $0.1 \, \rm{Z}_{\odot}$, shown by the red and blue curves in the bottom panel. 

For comparison, the vertical dotted lines show the minimum density of the cold neutral medium (CNM) predicted by the model of \citet{wolfire03}, if the CNM is in pressure equilibrium with the warm neutral medium (WNM). To determine this minimum density, \citet{wolfire03} calculate the thermal equilibrium temperature of the gas as a function of density, assuming that photoelectric heating from dust grains is balanced by radiative cooling from C\textsc{ii} and O\textsc{i}. Using this temperature-density relation, they determine the minimum pressure at which two stable ISM phases can exist in pressure equilibrium. This minimum pressure corresponds to the minimum density of the CNM, which is given by their equation 35:

\begin{equation}\label{wolfire_eqn}
n_{\rm{min}} \simeq 31 \frac{G_{0}^{\prime} (Z_{\rm{d}}^{\prime} / Z_{\rm{g}}^{\prime})}{1 + 3.1 (G_{0}^{\prime} Z_{\rm{d}}^{\prime} / \zeta_{\rm{t}}^{\prime})^{0.365}} \, \rm{cm}^{-3}, 
\end{equation}
where $G_{0}$ is the strength of the UV radiation field in units of the \citet{habing68} field, $Z_{\rm{d}}$ is the abundance of dust and polyaromatic hydrocarbons (PAHs), $Z_{\rm{g}}$ is the gas phase metallicity, and $\zeta_{\rm{t}}$ is the ionisation rate from cosmic rays and EUV/X-ray radiation. Primes indicate that these values have been normalised to their values in the local solar neighbourhood, for which \citet{wolfire03} take $G_{0} = 1.7$ and $\zeta_{\rm{t}} = 10^{-16} \, \rm{s}^{-1}$. 

At solar metallicity, the minimum CNM density predicted by \citet{wolfire03} coincides with the peak of the density distribution from our simulations, below which the distribution declines rapidly. Since this low-density tail is more extended in the simulations run with the full non-equilibrium chemical model, we find more cold gas below the minimum density of the \citet{wolfire03} model in this case than when we evolve the galaxy with equilibrium cooling rates. For example, at $Z = \rm{Z}_{\odot}$, $27.9$ per cent of the cold gas mass has a density $n < n_{\rm{min}}$ when we use non-equilibrium cooling rates, compared to $17.9$ per cent when we use cooling rates in chemical equilibrium. 

Note that, in the \citet{wolfire03} model, the CNM has a maximum temperature of $243 \, \rm{K}$, whereas we consider cold gas with $T < 10^{3} \, \rm{K}$, since this corresponds to the temperature threshold that we use in our star formation prescription. If we consider the density distribution of gas with $T < 243 \, \rm{K}$ in our simulations, we find that $2.4$ per cent of the gas mass has a density $n < n_{\rm{min}}$ in the non-equilibrium run at solar metallicity, compared to $0.6$ per cent in the equilibrium run. Thus our simulations are not in conflict with the predictions of \citet{wolfire03}. We continue to use $n_{\rm{min}}$ as a convenient metallicity- and radiation field-dependent reference point in our comparisons below. 

At lower metallicities, the peak of the density distribution of cold gas moves to higher densities. The minimum density predicted by \citet{wolfire03} also increases as the metallicity decreases, although it does not increase as quickly as in our simulations. We thus find less cold gas with $n < n_{\rm{min}}$ at lower metallicity. For example, at $Z = 0.1 \, \rm{Z}_{\odot}$, $14.2$ per cent of the cold ($T < 10^{3} \, \rm{K}$) gas mass has $n < n_{\rm{min}}$ when we use non-equilibrium cooling, compared to $6.5$ per cent when we use cooling rates in chemical equilibrium. 

\begin{figure*}
\centering
\mbox{
	\includegraphics[width=112mm]{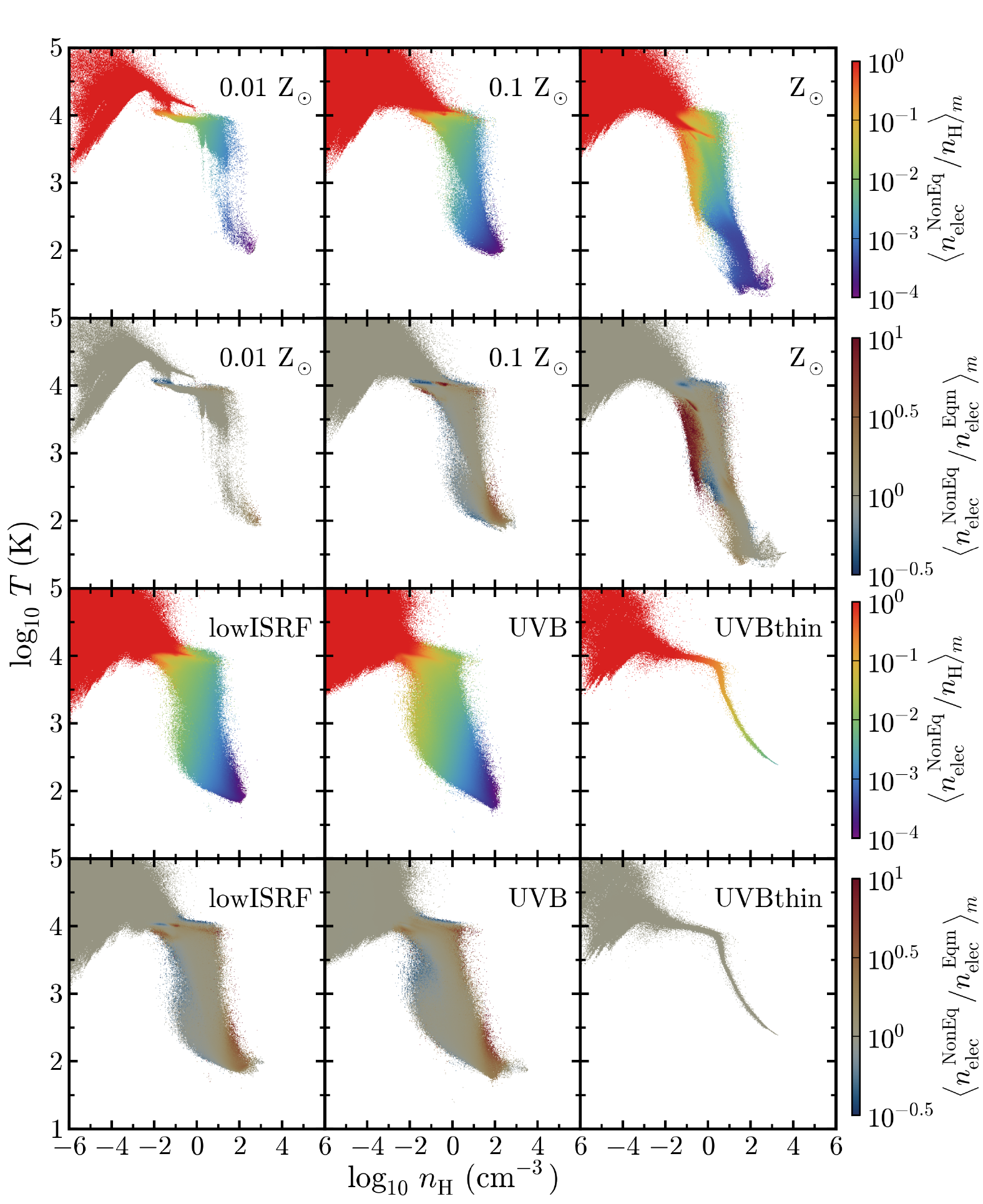}}
\caption{Temperature-density diagrams from simulations run with the full non-equilibrium chemical model at different metallicities \textit{(top two rows)} and for different radiation fields \textit{(bottom two rows; see Table~\ref{uv_table})}. The colour scale indicates the mass-weighted mean non-equilibrium electron abundance \textit{(first and third rows)} and the mass-weighted mean ratio between the non-equilibrium electron abundance and the electron abundance in equilibrium \textit{(second and fourth rows)}. We have averaged over snapshot outputs taken at intervals of $100 \, \rm{Myr}$. Red regions in the second and fourth rows show where electrons are enhanced with respect to equilibrium, and blue regions show where it is suppressed.}
\label{nonEqRatio_elec_fig}
\end{figure*}

\begin{figure*}
\centering
\mbox{
	\includegraphics[width=112mm]{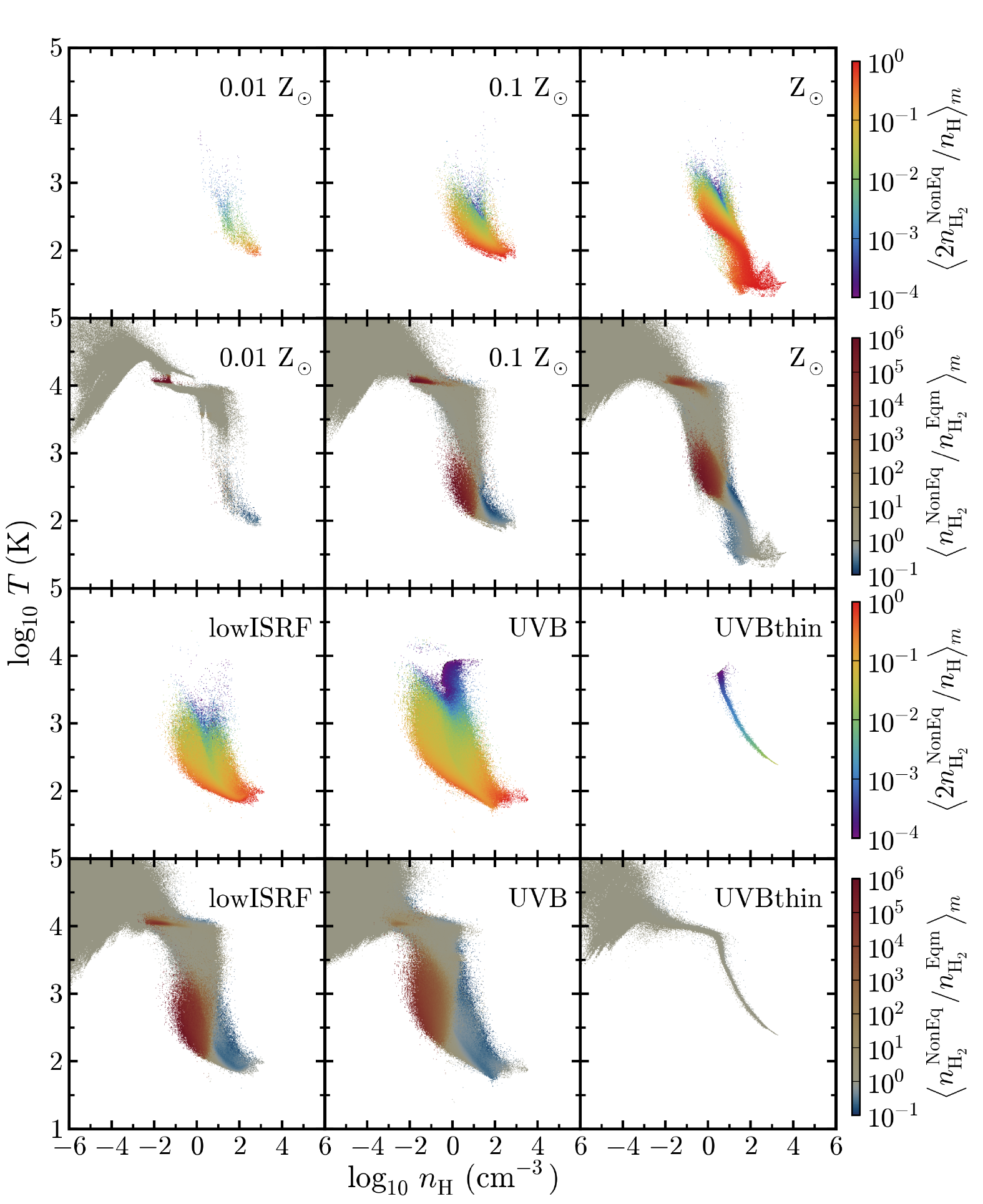}}
\caption{As Fig.~\ref{nonEqRatio_elec_fig}, but with the colour scale now indicating the mass-weighted mean non-equilibrium $\rm{H}_{2}$ abundance \textit{(first and third rows)} and the mass-weighted mean ratio between the non-equilibrium $\rm{H}_{2}$ abundance and the $\rm{H}_{2}$ abundance in equilibrium \textit{(second and fourth rows)}. We see that the H$_{2}$ abundance at $T < 10^{3} \, \rm{K}$ and $n_{\rm{H}} \sim 1 \, \rm{cm}^{-3}$ is strongly enhanced, by up to six orders of magnitude, at $Z \geq 0.1 \, \rm{Z}_{\odot}$. The resulting increase in H$_{2}$ cooling in this region explains why we see more cold gas at low densities in the one-dimensional density pdfs in Fig.~\ref{rho_pdf_fig} when we use non-equilibrium cooling compared to equilibrium cooling.}
\label{nonEqRatio_H2_fig}
\end{figure*}

In the presence of a weaker UV radiation field (i.e. lower $G_{0}$), the density distribution of cold gas moves to lower densities in our simulations, as does the minimum CNM density predicted by \citet{wolfire03}. We also see that the differences between non-equilibrium and equilibrium cooling become more pronounced in the presence of weaker UV radiation fields. For example, in the simulations run in the presence of the \citet{haardt01} UVB (blue curves in the bottom panel of Fig.~\ref{rho_pdf_fig}), $13.2$ per cent of the cold gas mass in the simulation run with non-equilibrium cooling has a density below $n_{\rm{min}}$ predicted by equation~\ref{wolfire_eqn}, compared to 0.8 per cent in the corresponding simulation evolved using cooling rates in chemical equilibrium. 

To understand why non-equilibrium cooling causes these differences in the temperature-density distribution, it is useful to explore in which regions of the temperature-density plane the chemical abundances are out of equilibrium. In Fig.~\ref{nonEqRatio_elec_fig} we show temperature-density diagrams with a colour scale indicating the mass-weighted mean electron abundance (first and third rows) and the mass-weighted mean ratio between the non-equilibrium electron abundance and the abundance of electrons in chemical equilibrium (second and fourth rows). Red and blue in the second and fourth rows indicate that the electron abundance is enhanced and reduced respectively, with respect to equilibrium. The top two rows and bottom two rows of Fig.~\ref{nonEqRatio_elec_fig} show variations in metallicity and radiation field respectively. 

At metallicities $Z \leq 0.1 \, \rm{Z}_{\odot}$ the electron abundance is close to equilibrium throughout most of the simulation. In the centre panel of the second row of Fig.~\ref{nonEqRatio_elec_fig} the electron abundance at densities $n_{\rm{H}} \sim 10^{2} \, \rm{cm}^{-3}$ is enhanced by a factor of a few compared to equilibrium, and cold gas (with $T < 10^{3} \, \rm{K}$) at densities $n_{\rm{H}} \sim 1 \, \rm{cm}^{-3}$ shows electron abundances that are reduced by $\sim 20 - 40$ per cent below equilibrium. The narrow region between the warm ionised and warm neutral phases at $T \sim 10^{4} \, \rm{K}$ also shows electron abundances that are out of equilibrium, by up to an order of magnitude. 

Similar trends are also seen in the presence of a weaker radiation field, in the left and centre panels of the bottom row of Fig.~\ref{nonEqRatio_elec_fig}. When we neglect self-shielding, in the right panels of the bottom two rows of Fig.~\ref{nonEqRatio_elec_fig}, the electron abundance is everywhere consistent with chemical equilibrium. 

At solar metallicity, in the right panels of the top two rows of Fig.~\ref{nonEqRatio_elec_fig}, there is also a region at $n_{\rm{H}} \sim 0.1 \, \rm{cm}^{-3}$ that extends from $T \sim 250 \, \rm{K}$ to $T \sim 4000 \, \rm{K}$ where the electron abundance is enhanced by about an order of magnitude. We find that there is no gas in this region of the temperature-density plane in the simulation evolved with cooling rates in chemical equilibrium. In the non-equilibrium simulation, the enhanced electron abundance increases the collisional excitation rate of ions (such as C\textsc{ii}) by electrons and thus increases the radiative cooling rate, allowing it to cool below the thermal equilibrium temperature corresponding to chemical equilibrium. 

In Fig.~\ref{nonEqRatio_H2_fig} we similarly show temperature-density diagrams with a colour scale indicating the non-equilibrium $\rm{H}_{2}$ abundance (first and third rows) and the ratio between the non-equilibrium and equilibrium $\rm{H}_{2}$ abundances (second and fourth rows), for different metallicities (top two rows) and different radiation fields (bottom two rows). 

At metallicities $Z \geq 0.1 \, \rm{Z}_{\odot}$, in the centre and right panels of the second row of Fig.~\ref{nonEqRatio_H2_fig}, and at lower radiation field strengths, in the left and centre panels of the bottom row of Fig.~\ref{nonEqRatio_H2_fig}, we see that gas with $T < 10^{3} \, \rm{K}$ and $n_{\rm{H}} \sim 1 \, \rm{cm}^{-3}$ has a very strongly enhanced $\rm{H}_{2}$ abundance, by up to six orders of magnitude. We find that these gas particles were previously in molecular clouds, with densities $\sim 10 - 100$ times their current value in the previous $\sim 5 \, \rm{Myr}$. These molecular clouds then became disrupted or destroyed, and the gas moved to lower densities. However, since it takes a finite time for the molecular hydrogen to be destroyed and reach a new equilibrium, they retain a high molecular fraction, resulting in a strong overabundance of H$_{2}$ with respect to equilibrium.

Similar non-equilibrium effects in the H$_{2}$ fraction were also found by \citet{dobbs08} in their simulations of spiral galaxies. However, it is possible that this enhancement in the H$_{2}$ abundance is sensitive to the resolution of our simulations. For example, this low-density gas may in reality be clumpy on scales smaller than we resolve, which would make it less well shielded from the photodissociating radiation than in our simulations. 

This region of enhanced H$_{2}$ in the temperature-density plane corresponds to the extended low-density tail that we saw in the density distributions of cold ($T < 10^{3} \, \rm{K}$) gas in Fig.~\ref{rho_pdf_fig}. The enhanced $\rm{H}_{2}$ abundance increases the cooling rate from $\rm{H}_{2}$, which allows this low-density gas to cool to lower temperatures. The enhanced H$_{2}$ abundances in outflowing gas that we saw in Fig.~\ref{H2_outflow_fig} also correspond primarily to this region of the temperature-density plane.

At higher densities, we also see blue regions in the second row of Fig.~\ref{nonEqRatio_H2_fig} at all metallicities, and in the left and centre panels of the bottom row, where the $\rm{H}_{2}$ abundance is reduced below its equilibrium value by up to an order of magnitude. We find that these gas particles were previously at lower densities and higher temperatures, and are now starting to become molecular. 

\citet{pelupessy09} also explored non-equilibrium H$_{2}$ chemistry in their hydrodynamic simulations of isolated galaxies, with metallicities up to $\rm{Z}_{\odot}$. They found H$_{2}$ fractions that were often far out of equilibrium in their simulations, in agreement with our results. In contrast, \citet{krumholz11} compared the time-dependent H$_{2}$ model from \citet{gnedin11} to the equilibrium H$_{2}$ model of \citet{krumholz08, krumholz09} and \citet{mckee10}, implemented in cosmological simulations, and they found good agreement between these two models at metallicities $\ga 0.01 \, \rm{Z}_{\odot}$. They therefore concluded that their equilibrium treatment of H$_{2}$ is sufficient above one per cent solar metallicity. However, their simulations used a lower resolution than we use. For example, the maximum resolution in their zoom-in cosmological simulations was $65 \, \rm{pc}$, whereas our simulations use a gravitational softening length of $3.1 \, \rm{pc}$. This difference in resolution may explain why we find non-equilibrium chemistry to be more important than \citet{krumholz11}. 

\section{Observable Line Emission}\label{line_emission}

To investigate the effects of metallicity, radiation field and non-equilibrium chemistry on observable diagnostics, we computed line emission maps for C\textsc{ii} and CO. We used the publicly available Monte-Carlo radiative transfer code \textsc{radmc-3d}\footnote{\url{http://www.ita.uni-heidelberg.de/~dullemond/software/radmc-3d/}} (version 0.38) to compute the line emission from these species by post-processing the output from our simulations. \textsc{radmc-3d} includes thermal emission from dust and line emission from user-specified species and transitions. We also include anisotropic scattering of continuum and line emission by dust grains, although in the current version of \textsc{radmc-3d} scattering of line emission does not include the corresponding doppler shift due to the relative motion of the dust, it only changes the direction of the radiation. 

We include two populations of dust grains, graphite and silicate, using opacities from the calculations of \citet{martin90}, who used the power-law grain size distribution of \citet{mathis77}. We use a dust-to-gas mass ratio of $2.4 \times 10^{-3} \, Z / \rm{Z}_{\odot}$ and $4.0 \times 10^{-3} \, Z / \rm{Z}_{\odot}$ for the graphite and silicate grains respectively, which we take from the `ISM' grain abundances used in the photoionisation code \textsc{cloudy}\footnote{\url{http://nublado.org/}} version $13.01$ \citep{ferland13}. 

The line emission from a given species depends on its level populations. However, unless we assume that these level populations are in Local Thermodynamic Equilibrium (LTE), they will also depend on the radiation field, including the emission lines themselves. A full, self-consistent non-LTE treatment can therefore be computationally expensive, as line emission from one gas cell can influence the level populations of its neighbours. Nevertheless, non-LTE effects can be important. \citet{duartecabral15} created synthetic maps of CO emission from their hydrodynamic simulations of spiral galaxies, and they compared maps created with and without assuming LTE. They found that, while the morphology of the CO emission was unaffected, the CO line intensity was generally overestimated when assuming LTE. 

\textsc{radmc-3d} includes a number of approximate methods to include non-LTE effects. We use the Large Velocity Gradient (LVG) method, which is also known as the Sobolev approximation \citep{sobolev57}. This method assumes that, after propagating for some distance, an emission line will be sufficiently Doppler shifted, due to motions of the gas, to propagate freely. We can thus calculate an escape probability for emitted photons based on the velocity gradient, which allows us to calculate the level populations from local quantities. A full description of the LVG method as implemented in \textsc{radmc-3d} can be found in \citet{shetty11}. 

With the LVG method, the non-LTE line emission from species $i$ depends on the density of $i$ and the species $j$ that can collisionally excite it, along with the gas temperature, gas velocity and the radiation field. Apart from the radiation field, we compute these quantities, and the densities of graphite and silicate grains, on a $4.0 \, \rm{kpc} \times 4.0 \, \rm{kpc} \times 4.0 \, \rm{kpc}$ Cartesian grid with a resolution of $10 \, \rm{pc}$ per cell. We calculate these quantities from the SPH particles in the snapshot outputs using SPH interpolation with the same Wendland C2 kernel with 100 SPH neighbours as was used in the simulations. We then compute the line and thermal dust emission, viewing the disc face-on, in 80 wavelength bins centred on the line and covering a velocity range of $\pm 40 \, \rm{km} \, \rm{s}^{-1}$. Finally, we repeat this without the emission line to create a map of the thermal dust emission only, and we subtract this from the full map to obtain the continuum-subtracted line emission. 

It is important to note that the C\textsc{ii} and CO emission presented in this section may be affected by the resolution of our simulations. In particular, both C\textsc{ii} and CO have a critical density of $\sim 10^{3} \, \rm{cm}^{-3}$. However, we saw in Fig.~\ref{T_rho_fig} that structures with such high densities are not well resolved in our simulations. Therefore, we might underestimate the C\textsc{ii} and CO emission from dense, unresolved structures. Additionally, high-resolution simulations of dense clouds find that most CO is concentrated in compact ($\sim 1 \, \rm{pc}$), high density ($\sim 10^{3} \, \rm{cm}^{-3}$) clumps and filaments \citep[e.g.][]{glover12}. 

\subsection{CII fine-structure line emission}

\begin{figure}
\centering
\mbox{
	\includegraphics[width=84mm]{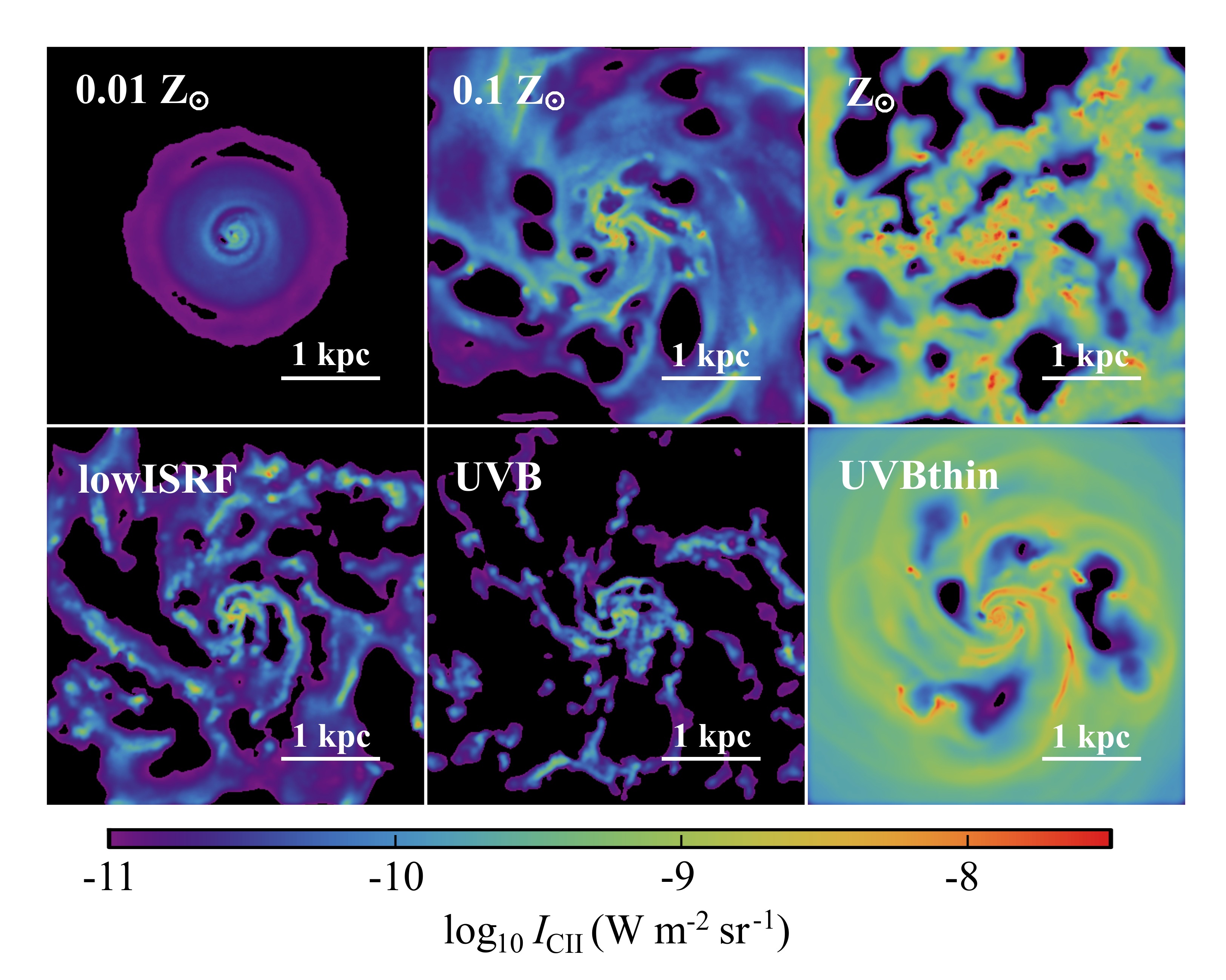}}
\vspace{-0.2 in}
\caption{C\textsc{ii} $158 \, \mu\rm{m}$ line emission from simulations evolved with the full non-equilibrium chemical model at different metallicities \textit{(top row)} and for different radiation fields \textit{(bottom row; see Table~\ref{uv_table})}. Each map shows the central region of the galaxy, $4.0 \, \rm{kpc}$ across, at $500 \, \rm{Myr}$, viewing the disc face on. C\textsc{ii} emission increases with increasing metallicity and increasing radiation field.}
\label{cii_emission_fig}
\end{figure}

The C\textsc{ii} fine-structure line at $158 \, \mu\rm{m}$ is an important coolant in neutral, atomic gas \citep[e.g.][]{richings14a}. This line is therefore commonly used as an observational tracer of the cooling properties and physical conditions of the neutral phases of the ISM \citep[e.g.][]{malhotra01, brauher08, graciacarpio11, kennicutt11, beirao12, croxall12}.

We calculate the emission from the C\textsc{ii} $158 \, \mu\rm{m}$ line using \textsc{radmc-3d}, as described above, with atomic data for the C\textsc{ii} ion taken from the \textsc{lamda} database\footnote{\url{http://home.strw.leidenuniv.nl/~moldata/}} \citep{schoier05}. These include excitation rates from collisions with ortho- and para-H$_{2}$ \citep{lique13, wiesenfeld14}, for which we assume an ortho-to-para ratio of 3:1, neutral H\textsc{i} \citep{barinovs05}, and electrons \citep{wilson02}. 

Fig.~\ref{cii_emission_fig} shows C\textsc{ii} line emission maps of the central region of each galaxy, $4.0 \, \rm{kpc}$ across, at $500 \, \rm{Myr}$, viewing the disc face on. The top row shows variations in metallicity, while the bottom row shows different UV radiation fields. These maps were computed from simulations evolved with the full chemical model, using the non-equilibrium abundances of C\textsc{ii} and the collision species (H$_{2}$, H\textsc{i} and electrons). 

Comparing the galaxies at different metallicities $Z$, we see that the total C\textsc{ii} line emission increases approximately linearly with $Z$. We might have expected such a linear relationship as increasing $Z$ will increase the amount of carbon in the gas. However, this simple picture is complicated by the fact that the densities and temperatures of the neutral gas are also affected by the metallicity, as we saw in Fig.~\ref{T_rho_fig}, which will affect the emissivities per C\textsc{ii} ion. A better way to explain this trend of C\textsc{ii} emission with metallicity is to note that, if the neutral gas is in thermal equilibrium, the cooling rate of the gas (which comes primarily from C\textsc{ii} and O\textsc{i} line emission) will balance the heating rate (which is mainly from photoelectric heating from dust grains). The photoelectric heating rate \citep{bakes94, wolfire95} scales with the abundance of dust grains, which we assume scales linearly with $Z$. Therefore, at higher metallicity, the photoelectric heating rate increases, and thus we need more C\textsc{ii} emission to achieve thermal balance. 

We also see very different morphologies of C\textsc{ii} emission at different metallicities. At $Z \geq 0.1 \, \rm{Z}_{\odot}$, we see C\textsc{ii} emission from dense clumps that are loosely arranged into spiral arms. In contrast, at $Z = 0.01 \, \rm{Z}_{\odot}$ we see an almost undisturbed disc, with C\textsc{ii} emission peaking in the centre. This reflects the different morphologies of the total gas distribution that we saw in Fig.~\ref{gasDensityFig}, which are due to the disruption of the disc by stellar feedback and the ability of the gas to cool to lower temperatures and higher densities at higher metallicities, as discussed in section~\ref{sf_section}. 

Comparing galaxies with different UV radiation fields at ten per cent solar metallicity (top centre, bottom left and bottom centre panels of Fig.~\ref{cii_emission_fig}), we see that the total C\textsc{ii} emission increases as the strength of the radiation field increases. This can also be understood as the photoelectric heating rate is higher in the presence of a stronger UV field, and thus more C\textsc{ii} emission is needed to balance the heating rate. However, the photoelectric heating rate does not increase linearly with radiation field strength, because dust grains become positively charged in the presence of a strong UV field, which reduces the photoelectric heating efficiency. 

In the run without self-shielding (bottom right panel), we find much stronger C\textsc{ii} emission, by nearly an order of magnitude compared to the corresponding simulation that includes self-shielding (bottom centre panel). This is because the dense gas is much warmer when self-shielding is ignored. For example, at $n_{\rm{H}} = 100 \, \rm{cm}^{-3}$, $T \sim 500 \, \rm{K}$ when self-shielding is not included, compared to $T \sim 50 - 100 \, \rm{K}$ when it is included (see Fig.~\ref{T_rho_fig}). Another way to see this is that, without self-shielding, the hydrogen-ionising radiation persists in the dense gas, so we have additional heating from photoionisation of hydrogen. We therefore need more C\textsc{ii} emission to achieve thermal balance. 

\begin{figure}
\centering
\mbox{
	\includegraphics[width=84mm]{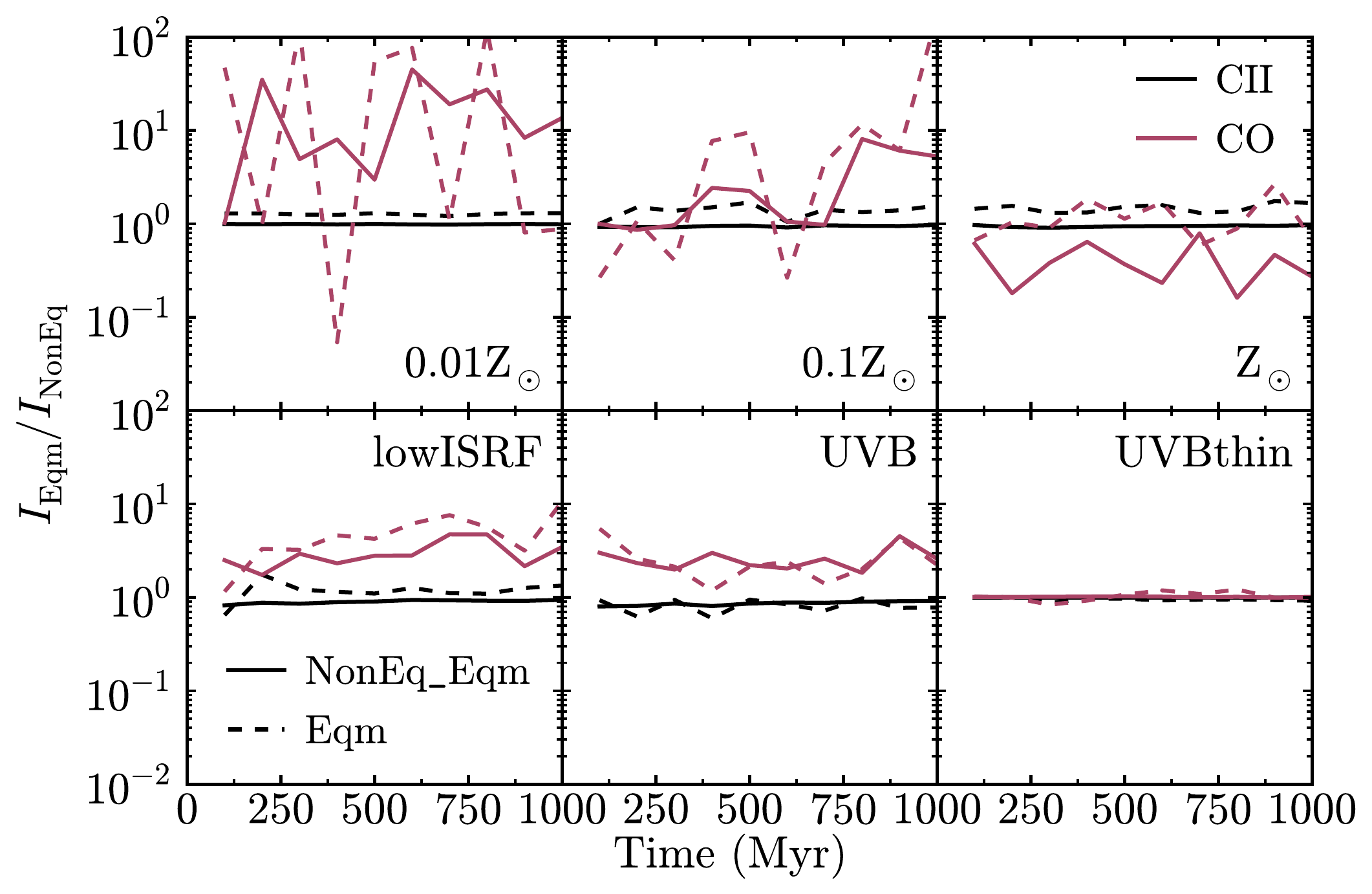}}
\caption{Ratio of average line intensity assuming equilibrium abundances, $I_{\rm{Eqm}}$, to that using non-equilibrium abundances, $I_{\rm{NonEq}}$, plotted against time, for different metallicities \textit{(top row)} and different UV radiation fields \textit{(bottom row; see Table~\ref{uv_table})}. We calculated $I_{\rm{Eqm}}$ from simulations evolved with the full non-equilibrium chemical model with abundances set to equilibrium in post-processing \textit{(NonEq\_Eqm; solid curves)}, and from simulations evolved in chemical equilibrium \textit{(Eqm; dashed curves)}. We show the C\textsc{ii} $158 \, \mu\rm{m}$ line \textit{(black curves)} and the CO $J = 1 - 0$ line \textit{(red curves)}.}
\label{line_noneq_fig}
\end{figure}

One caveat to note here is that, since we use a uniform UV radiation field, and we do not model local sources of radiation such as young stars, our simulations do not include H\textsc{ii} regions and PDRs. However, these are often strong sources of C\textsc{ii} emission. Therefore, it is possible that we are underestimating the C\textsc{ii} emission. 

We saw in section~\ref{phase_structure} that evolving a galaxy using cooling rates in non-equilibrium can affect the distribution of gas densities and temperatures, compared to using cooling rates in chemical equilibrium. We also found that some chemical species (particularly the free electrons and H$_{2}$) can be far out of equilibrium at certain densities and temperatures. These two effects could potentially have an impact on the observable line emission from individual species. 

To investigate the impact of non-equilibrium chemistry on the line emission maps, we also computed these from equilibrium abundances in two ways. Firstly, we used the simulations evolved with the full non-equilibrium chemical model and set the chemical abundances of each gas particle to chemical equilibrium (`NonEq\_Eqm'). This shows how non-equilibrium abundances affect the line emission. Secondly, we used the simulations that were evolved using cooling rates in chemical equilibrium (`Eqm'). This shows how the different distributions of gas density and temperature, due to using non-equilibrium cooling, affect the line emission. 

Fig.~\ref{line_noneq_fig} shows the ratio of the average line intensity assuming equilibrium abundances to that using non-equilibrium abundances, plotted against time. Solid and dashed curves show `NonEq\_Eqm' and `Eqm' respectively. C\textsc{ii} line emission is shown by the black curves, while the red curves show CO line emission, which we will discuss in section~\ref{co_emission_section}. The top row shows simulations run at different metallicities, and the bottom row shows for different UV radiation fields. 

The black solid curves, for C\textsc{ii} `NonEq\_Eqm', are all very close to unity in all panels. This tells us that the abundances of C\textsc{ii} and those species that collisionally excite it are close to equilibrium in C\textsc{ii}-emitting gas for the non-equilibrium runs. The black dashed curves, from the simulations evolved in chemical equilibrium, also remain close to unity, except in runs at ten per cent solar and solar metallicity in the presence of the \citet{black87} ISRF (top centre and top right panels), where the equilibrium emission is $\sim 50$ per cent higher. 

\subsection{CO line emission}\label{co_emission_section}

\begin{figure}
\centering
\mbox{
	\includegraphics[width=84mm]{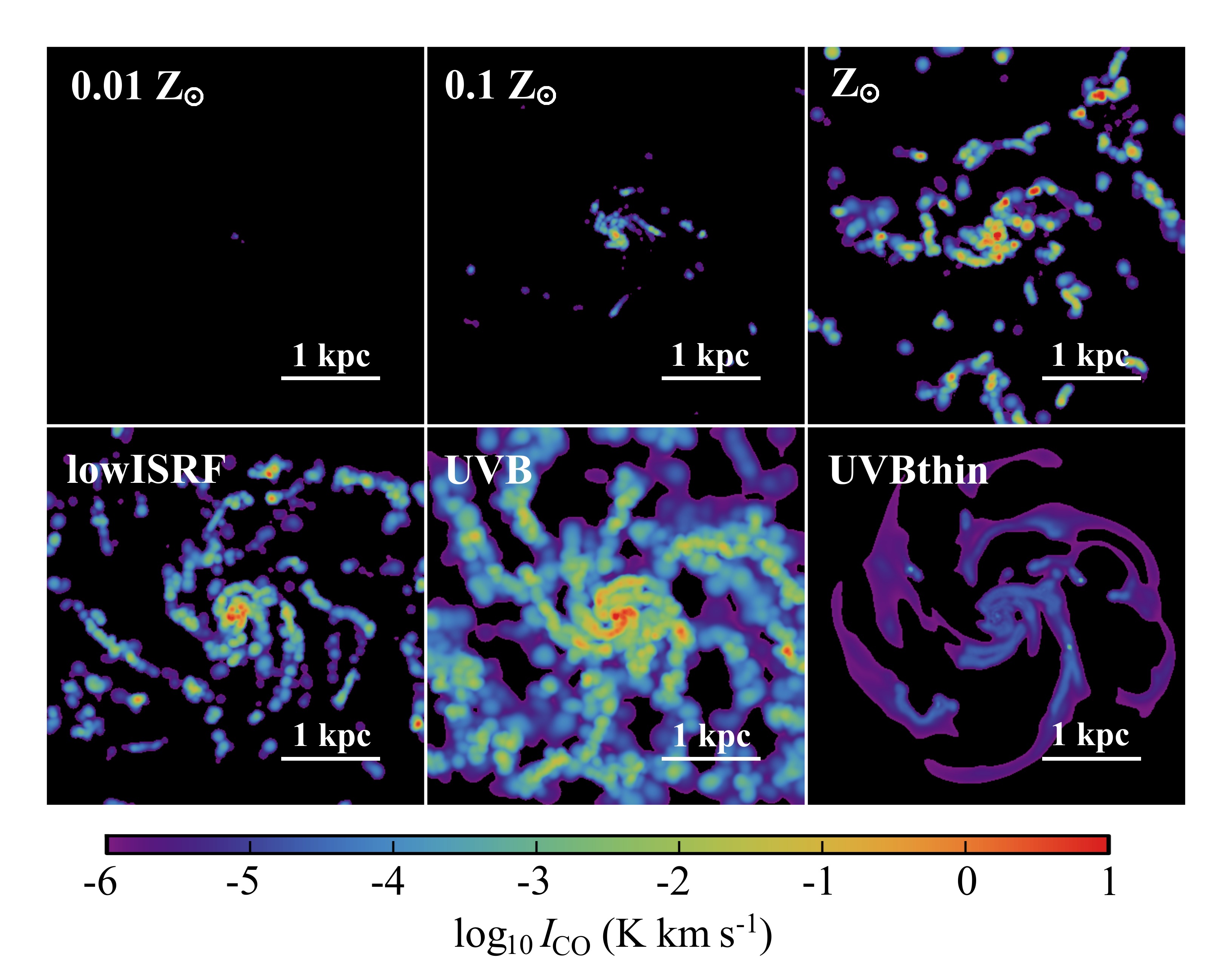}}
\vspace{-0.2 in}
\caption{As Fig.~\ref{cii_emission_fig}, but for CO $J = 1 - 0$ line emission at $2.6 \, \rm{mm}$. CO emission increases with increasing metallicity and decreasing UV radiation field.}
\label{co_emission_fig}
\end{figure}

Molecular hydrogen is difficult to observe directly in cold, molecular gas, because the lowest rovibrational levels of the H$_{2}$ molecule are difficult to excite at the temperatures typical of molecular clouds ($\sim 10 \, \rm{K}$). For example, the lowest rotational transition in the ground vibrational state, $0 - 0 \, S(0)$, has an excitation energy $E / k_{\rm{B}} = 510 \, \rm{K}$. However, one of the next most abundant molecules after H$_{2}$ is CO, which can be observed at much lower temperatures than H$_{2}$. For example, the lowest rotational transition of CO, $J = 1 - 0$, has an excitation energy of $5.53 \, \rm{K}$. CO emission is therefore commonly used to map cold molecular gas \citep[e.g.][]{helfer03, kuno07, bolatto08, leroy09}. However, to determine the H$_{2}$ content from CO emission alone, we need to know the conversion factor, $X_{\rm{CO}}$, between CO emission and H$_{2}$ column density (see \citealt{bolatto13} for a recent review). The $X_{\rm{CO}}$ factor is commonly defined as: 

\begin{equation}\label{co_equation}
X_{\rm{CO}} = \frac{N_{\rm{H_{2}}}}{I_{\rm{CO}}} \, \rm{cm}^{-2} (\rm{K} \, \rm{km} \, \rm{s}^{-1})^{-1}, 
\end{equation}
where $N_{\rm{H_{2}}}$ is the H$_{2}$ column density and $I_{\rm{CO}}$ is the velocity-integrated intensity of the CO $J = 1 - 0$ line. 

In this section we present the emission from the CO $J = 1 - 0$ line, at a wavelength of $2.6 \, \rm{mm}$, in our simulations, computed using \textsc{radmc-3d}. We use molecular CO data from the LAMDA database, including collisional excitation by ortho- and para-H$_{2}$ \citep{yang10}, for which we assume an ortho-to-para ratio of 3:1. 

We show velocity-integrated line emission maps for the $J = 1 - 0$ line in Fig.~\ref{co_emission_fig} for different metallicities (top row) and different radiation fields (bottom row). In the galaxy with the lowest metallicity ($0.01 \, \rm{Z}_{\odot}$; top left panel of Fig.~\ref{co_emission_fig}), and in the run without self-shielding (bottom right panel), the CO intensity is very low, and would not be detectable. For comparison, for the CO observations of Local Group galaxies in \citet{leroy11}, the 3$\sigma$ CO intensity threshold for the Small Magellanic Cloud is $0.25 \, \rm{K} \, \rm{km} \, \rm{s}^{-1}$. 

We see very weak CO emission in the lowest-metallicity run because there is less carbon and oxygen available to form CO. Additionally, there is less dust shielding at lower metallicity, which is needed to prevent the destruction of CO by photodissociation. Similarly, in the simulations run without self-shielding, we see little CO emission because photodestruction of CO is always efficient. 

In the remaining galaxies, we see that the CO emission is concentrated in dense clumps along the spiral arms. Comparing panels in the top row, the CO intensity increases with increasing metallicity, while in the bottom left and bottom centre panels, the CO intensity increases with decreasing radiation field strength. 

The ratio of the average CO line intensity assuming equilibrium to that using non-equilibrium abundances is shown by the red curves in Fig.~\ref{line_noneq_fig}. Non-equilibrium chemistry has a greater effect on line emission from CO than for C\textsc{ii}. For example, in the simulations run in the presence of ten per cent of the \citet{black87} ISRF at ten per cent solar metallicity (lowISRF, bottom left panel), the average CO line intensity computed in equilibrium is higher by a factor of $\sim 4$ than in non-equilibrium. We also see very large relative differences between equilibrium and non-equilibrium CO emission at the lowest metallicity ($0.01 \, \rm{Z}_{\odot}$; top left panel), although we saw in Fig.~\ref{co_emission_fig} that the CO emission is very weak in this run, and would not be detectable in typical CO surveys. 

These non-equilibrium trends in CO emission are not always in the same direction. For example, at solar metallicity (top right panel of Fig.~\ref{line_noneq_fig}), the average CO line intensity calculated by setting the abundances to equilibrium in post-processing (`NonEq\_Eqm', solid red curve) is lower than in non-equilibrium by a factor of $\sim 2$. This is opposite to the trend that we saw in the lowISRF run. We find that there are two competing non-equilibrium effects that act on the CO abundance, and hence on the CO intensity. Firstly, gas that is forming into molecular clouds takes a finite time to form CO. Such gas is underabundant in CO with respect to equilibrium. Secondly, when an existing molecular cloud is disrupted or destroyed, it takes a finite time for the CO in this gas to be destroyed. Hence, CO is overabundant in such gas. 

Therefore, the effect of non-equilibrium chemistry on CO emission is not a simple one, as there are two competing effects, which depend crucially on the thermal history of the gas. Thus, non-equilibrium abundances can either increase or decrease the CO intensity. 

\begin{figure}
\centering
\mbox{
	\includegraphics[width=84mm]{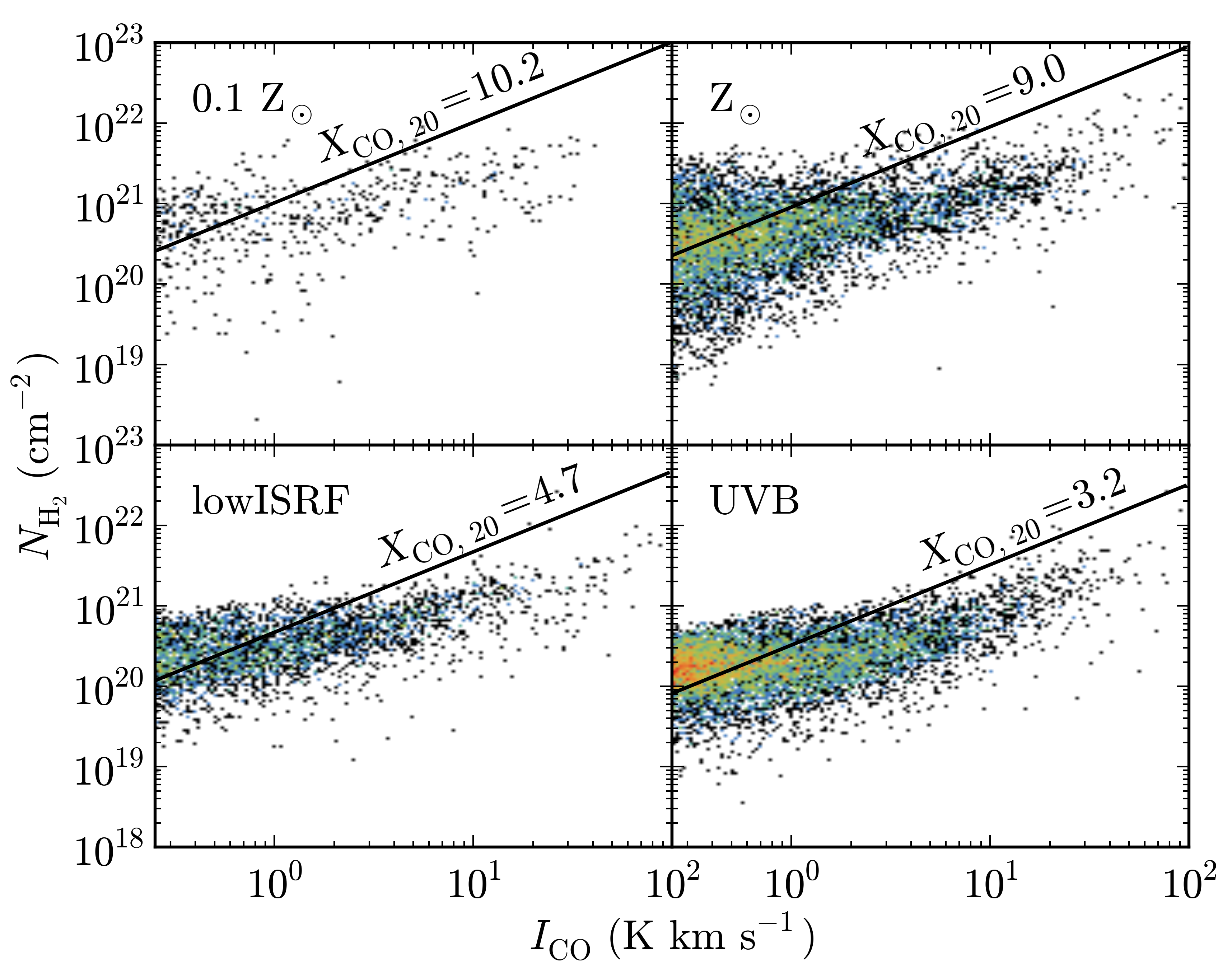}}
\caption{Relation between H$_{2}$ column density, $N_{\rm{H_{2}}}$, and velocity-integrated intensity of the CO $J = 1 - 0$ line, $I_{\rm{CO}}$, measured in pixels $10 \, \rm{pc}$ across that span the central $4 \, \rm{kpc}$ of the galaxy, viewing the disc face-on. We show simulations run with the full non-equilibrium chemical model at different metallicities \textit{(top row)} and for different radiation fields \textit{(bottom row; see Table~\ref{uv_table})}. We include ten snapshots from each simulation, taken at intervals of $100 \, \rm{Myr}$. Black lines show the linear $N_{\rm{H2}} - I_{\rm{CO}}$ relation obtained using the mean $X_{\rm{CO}}$ factor from the given simulation, averaged over pixels with $I_{\rm{CO}} > 0.25 \, \rm{K} \, \rm{km} \, \rm{s}^{-1}$. We express this mean conversion factor as $X_{\rm{CO}, \, 20} = X_{\rm{CO}} / (10^{20} \, \rm{cm}^{-2} \, (\rm{K} \, \rm{km} \, \rm{s}^{-1})^{-1})$.}
\label{xco_fig}
\end{figure}

In Fig.~\ref{xco_fig} we plot the H$_{2}$ column density of each pixel versus the velocity-integrated intensity of the CO $J = 1 - 0$ line. Each pixel in the emission line maps is $10 \, \rm{pc}$ across, and the maps cover the central $4 \, \rm{kpc}$ of the disc. We include ten snapshot outputs from each simulation, taken at intervals of $100 \, \rm{Myr}$. Many of the pixels have very low CO intensities that would be undetectable in a realistic survey of CO emission in extragalactic sources. We therefore only include pixels with a CO intensity $I_{\rm{CO}} > 0.25 \, \rm{K} \, \rm{km} \, \rm{s}^{-1}$, which corresponds to the 3$\sigma$ intensity threshold used by \citet{leroy11} for the Small Magellanic Cloud. 

We show the $N_{\rm{H_{2}}} - I_{\rm{CO}}$ relation from simulations run with the full non-equilibrium chemical model for different metallicities in the top row, and for different UV radiation fields in the bottom row. We do not include the simulations run at one per cent solar metallicity or run without self-shielding, as these do not show detectable CO emission. The black line in each panel shows the linear relation between $N_{\rm{H_{2}}}$ and $I_{\rm{CO}}$ using the mean $X_{\rm{CO}}$ factor measured in the simulation, averaged over pixels with $I_{\rm{CO}} > 0.25 \, \rm{K} \, \rm{km} \, \rm{s}^{-1}$. 

In Fig.~\ref{xco_fig}, we see that the mean $X_{\rm{CO}}$ factor increases with increasing radiation field strength, $G_{0}$. For comparison, measurements of the $X_{\rm{CO}}$ factor in molecular clouds in the Milky Way, using various different methods (Virial mass, CO isotopologues, dust extinction or emission, diffuse gamma-ray emission), find $X_{\rm{CO}, \, 20} = X_{\rm{CO}} / (10^{20} \, \rm{cm}^{-2} \, (\rm{K} \, \rm{km} \, \rm{s}^{-1})^{-1}) \sim 2 - 4$ (e.g. \citealt{bolatto13} and references therein). This trend with $G_{0}$ may arise because, in our simulations with different radiation fields (at $0.1 \, \rm{Z}_{\odot}$), the dust extinction only extends up to $A_{\rm{v}} \sim 1$. CO is therefore not fully shielded from photodissociation by dust in these simulations, whereas the molecular hydrogen can still become fully shielded, by self-shielding. Therefore, as $G_{0}$ increases, the CO intensity is suppressed more than the H$_{2}$ column density, and thus $X_{\rm{CO}}$ increases. 

In Fig.~\ref{xco_fig}, there is almost no dependence of $X_{\rm{CO}}$ on metallicity. This lack of dependence on metallicity may seem surprising, as observations find that $X_{\rm{CO}}$ decreases with increasing metallicity \citep[e.g.][]{israel97, leroy11}. However, this apparent discrepancy is likely because our simulations at ten per cent solar metallicity only probe regions with dust extinctions $A_{\rm{v}} \la 1$, while the solar metallicity runs extend up to $A_{\rm{v}} \sim 10$. 

To understand the trends of the $X_{\rm{CO}}$ factor at different $A_{\rm{v}}$, we can look at the model of \citet{feldmann12}. They construct a model for $X_{\rm{CO}}$ as a function of metallicity, $Z$, radiation field, $U$, and dust extinction, $A_{\rm{v}}$. They use H$_{2}$ and CO abundances from small-scale MHD simulations of the turbulent ISM from \citet{glover11}, which included a treatment for the non-equilibrium chemistry of H$_{2}$ and CO, although \citet{feldmann12} assume photodissociation equilibrium to determine the dependence of the CO abundance on $U$, as most of the simulations of \citet{glover11} were run for only one radiation field. \citet{feldmann12} compute the CO line emission based on an escape probability formalism, assuming that the level populations of the CO molecule are in LTE and using an assumption for the CO line width (either a constant line width or a virial scaling). 

\citet{feldmann12} show the dependence of $X_{\rm{CO}}$ on $A_{\rm{v}}$ in their model for various $Z$ and $U$ (see their fig. 2). They show that $X_{\rm{CO}}$ reaches a minimum at $A_{\rm{v}} \sim 2 - 10$, which approximately corresponds to where the CO line becomes optically thick. At $A_{\rm{v}}$ below this minimum (where the CO line is optically thin), they show that, at fixed $A_{\rm{v}}$, $X_{\rm{CO}}$ increases with $U$ (in agreement with our simulations), but is independent of $Z$. 

In our simulations at solar metallicity, we probe regions up to $A_{\rm{v}} \sim 10$. However, all our simulations at lower metallicities only extend up to $A_{\rm{v}} \sim 1$. Therefore, it is likely that we do not recover the observed trends of $X_{\rm{CO}}$ with metallicity because we do not cover a range of metallicities in the optically thick regime. 

\begin{figure}
\centering
\mbox{
	\includegraphics[width=84mm]{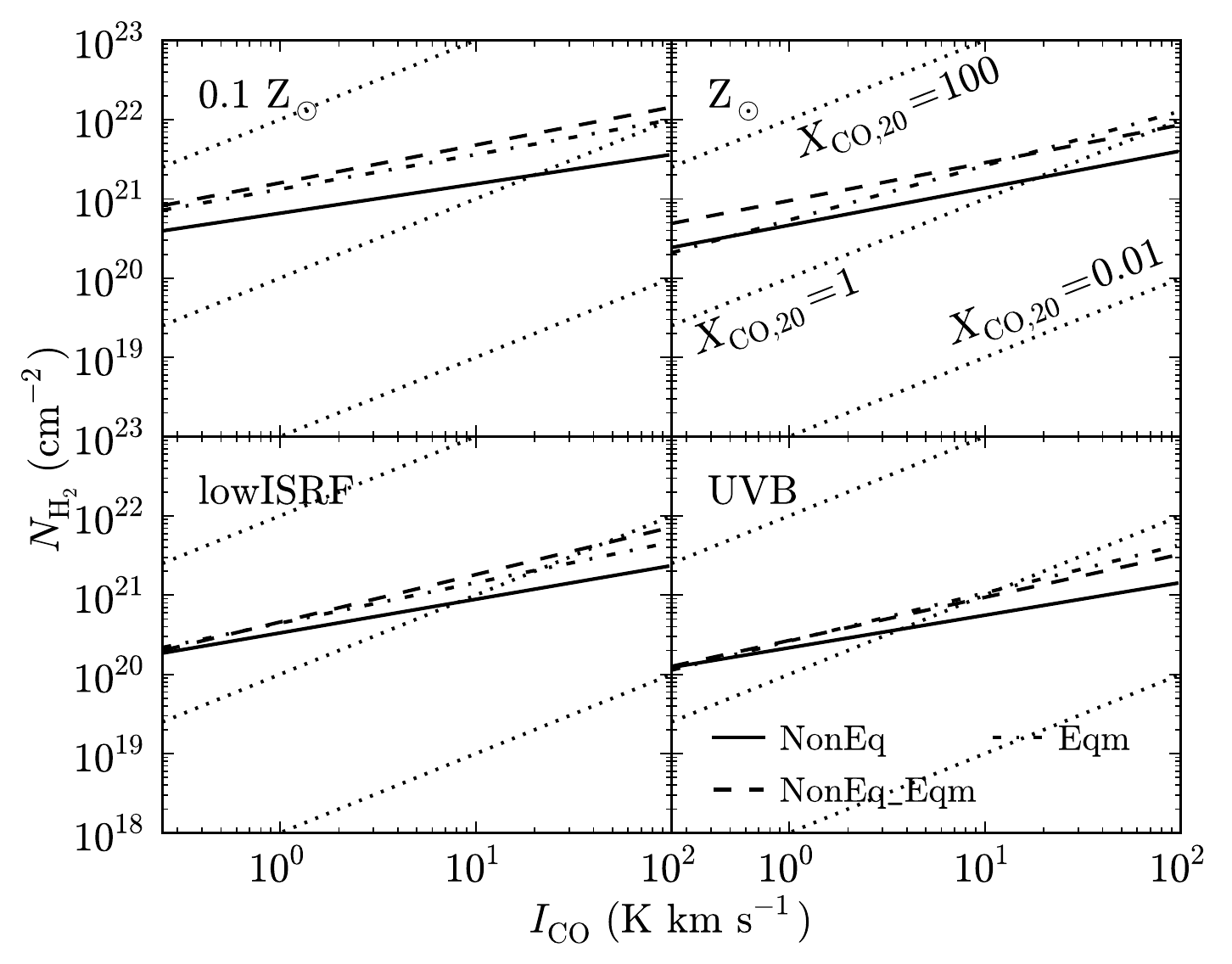}}
\caption{Best-fit power law relation between H$_{2}$ column density, $N_{\rm{H_{2}}}$, and velocity-integrated intensity of the CO $J = 1 - 0$ line, $I_{\rm{CO}}$, at different metallicities \textit{(top row)} and for different radiation fields \textit{(bottom row; see Table~\ref{uv_table})}. We use maps of $N_{\rm{H_{2}}}$ and $I_{\rm{CO}}$ computed from simulations evolved with the full non-equilibrium chemical model using non-equilibrium abundances \textit{(NonEq; solid curves)} and with abundances set to equilibrium in post-processing \textit{(NonEq\_Eqm; dashed curves)}, and from simulations evolved in chemical equilibrium \textit{(Eqm; dot-dashed curves)}. Dotted curves indicate lines of constant $X_{\rm{CO}}$, as indicated in the top right panel. Each relation was fit to pixels with $I_{\rm{CO}} > 0.25 \, \rm{K} \, \rm{km} \, \rm{s}^{-1}$.} 
\label{xco_noneq_fig}
\end{figure}

The $X_{\rm{CO}}$ conversion factor between CO intensity and H$_{2}$ column density, as defined in equation~\ref{co_equation}, may also be affected by non-equilibrium chemistry, which affects both the CO emission (as seen in Fig.~\ref{line_noneq_fig}) and the abundance of H$_{2}$ (as seen in Fig.~\ref{nonEqRatio_H2_fig}). Fig.~\ref{xco_noneq_fig} compares the best-fit power-law relation between $N_{\rm{H_{2}}}$ and $I_{\rm{CO}}$ from the non-equilibrium simulations (`NonEq'; solid curves) to those from the same simulations with abundances set to equilibrium in post-processing (`NonEq\_Eqm'; dashed curves), and from simulations evolved in chemical equilibrium (`Eqm'; dot-dashed curves). Dotted curves indicate lines of constant $X_{\rm{CO}}$. The top and bottom rows show different metallicities and UV radiation fields respectively. Each curve was fit to pixels with $I_{\rm{CO}} > 0.25 \, \rm{K} \, \rm{km} \, \rm{s}^{-1}$. 

The best-fit relations in Fig.~\ref{xco_noneq_fig} have power-law slopes of $\sim 0.4-0.6$. This is flatter than the linear relation that we would expect for a constant $X_{\rm{CO}}$ factor, and indicates that $X_{\rm{CO}}$ decreases as $I_{\rm{CO}}$ increases. 

The $N_{\rm{H_{2}}} - I_{\rm{CO}}$ relations calculated using non-equilibrium abundances (solid curves) are lower than those calculated in equilibrium. Table~\ref{xco_table} summarises the mean $X_{\rm{CO}}$ factors from each simulation, averaged over pixels with $I_{\rm{CO}} > 0.25 \, \rm{K} \, \rm{km} \, \rm{s}^{-1}$. 

\begin{table}
\centering
\begin{minipage}{84mm}
\caption{Mean $X_{\rm{CO}}$ factor, averaged over pixels with a velocity-integrated CO line intensity $I_{\rm{CO}} > 0.25 \, \rm{K} \, \rm{km} \, \rm{s}^{-1}$.}
\centering
\begin{tabular}{lccc}
\hline
 & \multicolumn{3}{c}{$X_{\rm{CO}} / 10^{20} \rm{cm}^{-2} (\rm{K} \, \rm{km} \, \rm{s}^{-1})^{-1}$} \\
Simulation & NonEq & NonEq\_Eqm & Eqm \\
\hline
ref & $10.2$ & $23.9$ & $23.0$ \\ 
hiZ & $9.0$ & $23.4$ & $13.6$ \\ 
lowISRF & $4.7$ & $5.9$ & $6.6$ \\ 
UVB & $3.2$ & $3.5$ & $3.3$ \\ 
\hline
\vspace{-0.1in}
\label{xco_table}
\end{tabular}
\end{minipage}
\end{table}

In the presence of the \citet{black87} ISRF (ref and hiZ), we find that the mean $X_{\rm{CO}}$ factor decreases by a factor $\sim 2.3$ when we use non-equilibrium abundances. At lower radiation field strengths (lowISRF and UVB), the effect of non-equilibrium chemistry on the mean $X_{\rm{CO}}$ factor is much smaller (e.g. $10$ per cent in the UVB runs). Nevertheless, we saw in Fig.~\ref{line_noneq_fig} that the mean CO intensity for weaker radiation fields is lower, by a factor of $\sim 4$, when we use non-equilibrium abundances than in equilibrium. We also find fewer pixels with $I_{\rm{CO}}$ above the detection threshold of $0.25 \, \rm{K} \, \rm{km} \, \rm{s}^{-1}$ when we use non-equilibrium abundances in these examples. 

If we find fewer pixels with detectable CO emission when we use non-equilibrium abundances, we might also expect this to have an impact on the fraction of CO-dark molecular gas, i.e. molecular hydrogen that is not traced by CO emission \citep[e.g.][]{tielens85, vandishoeck88, wolfire10, smith14}. We calculated the CO dark fraction in the simulations ref, hiZ, lowISRF and UVB. This is the mass of H$_{2}$ in pixels with $I_{\rm{CO}} < 0.25 \, \rm{K} \, \rm{km} \, \rm{s}^{-1}$ divided by the mass of H$_{2}$ in all pixels of the CO map. For each simulation, we averaged this CO dark fraction over ten snapshots taken at intervals of 100 Myr. When we use H$_{2}$ abundances and CO maps in non-equilibrium, the CO dark fraction is $0.76 - 0.86$ in these four simulations, with no strong dependence on metallicity or UV radiation field. However, when we set the abundances to chemical equilibrium, the CO dark fraction in the UVB and lowISRF runs decreases to 0.63 and 0.59, respectively. In the ref and hiZ runs, the effect of non-equilibrium chemistry is weaker, as the CO dark fraction only changes by ten per cent when we use equilibrium abundances. 

\citet{wolfire10} used theoretical models of molecular clouds to determine the CO dark fraction for different cloud masses and UV radiation fields. They found an almost constant CO dark fraction of 0.3, although they only considered individual clouds, whereas we compute the CO dark fraction over the disc of the galaxy, including diffuse gas. \citet{smith14} simulated four Milky Way-type galaxies with different gas surface densities and UV radiation fields. They found CO dark fractions of $0.4 - 0.85$. The highest values that they found are consistent with our simulations, although they also found that increasing the strength of the UV radiation field by a factor of 10 increases the CO dark fraction by a factor of $\approx 1.5-2$. We do not find any strong dependence on UV radiation field in our simulations. 

\section{Conclusions}\label{conclusions}

We have run a series of hydrodynamic simulations of isolated galaxies with a virial mass $M_{200, \rm{crit}} = 10^{11} \, \rm{M}_{\odot}$ and stellar mass $10^{9} \, \rm{M}_{\odot}$. The models use a resolution of $750 \, \rm{M}_{\odot}$ per gas particle and a gravitational force resolution of $3.1 \, \rm{pc}$, and were run with a modified version of the SPH code \textsc{gadget}3. We included a treatment for the full non-equilibrium chemical evolution of ions and molecules (157 species in total), along with gas cooling rates computed self-consistently from these non-equilibrium abundances \citep{richings14a, richings14b}, and we compared these to simulations evolved using cooling rates in chemical equilibrium. 

Our simulations were run at a fixed metallicity and in the presence of a uniform UV radiation field, with a local prescription for self-shielding by gas and dust. We covered a wide range of metallicities ($0.01 \, \rm{Z}_{\odot}$, $0.1 \, \rm{Z}_{\odot}$ and $\rm{Z}_{\odot}$), and different UV radiation fields that span nearly three orders of magnitude in H\textsc{i} photoionisation rate (the \citet{black87} ISRF, ten per cent of the \citet{black87} ISRF, and the redshift zero UVB of \citet{haardt01}; see Table~\ref{uv_table}), and we also repeated the runs with the \citet{haardt01} UVB without self-shielding. 

Our goal was to investigate the effects of metallicity, radiation field and non-equilibrium chemistry, which have all been demonstrated to affect gas cooling rates, in simulations of galaxy evolution. There are two aspects to the impact of these effects. Firstly, how the changes in gas cooling rates affect the evolution of the galaxy, and secondly, how observable tracers of individual chemical species are affected. 

Our main results are as follows: 

\begin{enumerate}
\item In simulations at higher metallicity, and for weaker UV radiation fields, gas can more easily cool to a cold ($T \sim 100 \, \rm{K}$), star forming ISM phase, due to increased metal-line cooling and reduced UV heating respectively. We thus find higher star formation rates in these cases, by two orders of magnitude and a factor $\sim 3$ for the different metallicities and different radiation fields, respectively, that we consider here (Fig.~\ref{SFH}). In particular, the gas surface density threshold below which star formation is cut off decreases with increasing metallicity and decreasing UV radiation field (Fig.~\ref{KS}), as predicted by \citet{schaye04}. 
\item We find higher mass outflow rates at higher metallicity (by two orders of magnitude) and for weaker radiation fields (by a factor $\sim 3$), due to the higher star formation rates (Fig.~\ref{radial_outflows_fig}). However, the average mass loading factor (i.e. the ratio of outflow to star formation rates), measured at $0.2 R_{200, \rm{crit}} = 19 \, \rm{kpc}$, is $\sim 10$, regardless of metallicity or radiation field. 
\item The mass loading factor, $\beta$, measured $1 \, \rm{kpc}$ above and below the disc decreases with increasing gas surface density, $\Sigma_{\rm{gas}}$, following approximately a power-law, $\beta \propto \Sigma_{\rm{gas}}^{-2}$ (Fig.~\ref{mass_loading_fig}). At fixed $\Sigma_{\rm{gas}}$, the mass loading factor increases with decreasing metallicity and increasing radiation field strength. 
\item Non-equilibrium cooling does not strongly affect the total star formation rate of the galaxy (Fig.~\ref{SFH}). The initial rise in mass outflow rate in the first $\sim 200 \, \rm{Myr}$ of the simulation is sometimes more gradual when we use equilibrium cooling, compared to the non-equilibrium runs (Fig.~\ref{radial_outflows_fig}). However, apart from this initial difference, non-equilibrium cooling does not strongly affect the outflow properties. 
\item Non-equilibrium chemistry does have a large effect on the chemical make-up of outflowing gas (Fig.~\ref{H2_outflow_fig}). For example, in our reference run (black curves in Fig.~\ref{H2_outflow_fig}), we find on average $600 \, \rm{M}_{\odot}$ of H$_{2}$ outflowing with a vertical velocity of $> 50 \, \rm{km} \, \rm{s}^{-1}$ if we use non-equilibrium abundances, compared to $30 \, \rm{M}_{\odot}$ if we assume chemical equilibrium. Non-equilibrium chemistry therefore enhances the mass of outflowing H$_{2}$ by a factor $\sim 20$ in this example. This has important implications for modelling molecular outflows in hydrodynamic simulations of galaxies. 
\item We investigated where in the temperature-density plane molecular hydrogen is out of equilibrium (Fig.~\ref{nonEqRatio_H2_fig}). H$_{2}$ can be enhanced, by up to six orders of magnitude, in gas that was previously in molecular clouds at higher densities but has since been disrupted. We also find regions, around $n_{\rm{H}} \sim 10 - 100 \, \rm{cm}^{-3}$, where H$_{2}$ is underabundant, by up to an order of magnitude. This is due to gas that is starting to form molecular clouds, but has not yet had enough time to fully form H$_{2}$. 
\item Using the publicly available Monte-Carlo radiative transfer code \textsc{radmc-3d}\footnote{\url{http://www.ita.uni-heidelberg.de/~dullemond/software/radmc-3d/}}, we performed radiative transfer calculations on our simulations in post-processing to compute the line emission from C\textsc{ii} and CO. C\textsc{ii} emission from the $158 \, \mu\rm{m}$ line is stronger at higher metallicity and for stronger radiation fields (Fig.~\ref{cii_emission_fig}), while CO emission from the $J = 1 - 0$ line is stronger at higher metallicity and for weaker radiation fields (Fig.~\ref{co_emission_fig}). 
\item C\textsc{ii} emission is generally unaffected by non-equilibrium chemistry, whereas CO emission is affected by a factor of $\sim 2 - 4$ (Fig.~\ref{line_noneq_fig}). However, the CO emission can be either higher or lower in non-equilibrium, since, similarly to H$_{2}$, the CO abundance can be either enhanced or suppressed. This also affects the mean $X_{\rm{CO}}$ conversion factor between CO line intensity and H$_{2}$ column density (equation~\ref{co_equation}) that we measure in the simulations, by up to a factor $\sim 2.3$ (Table.~\ref{xco_table}). 
\item Non-equilibrium chemistry also affects the fraction of CO-dark molecular gas \citep[e.g.][]{tielens85, vandishoeck88, wolfire10, smith14}, i.e. the fraction of molecular hydrogen that is not traced by observable CO emission. For example, in our `lowISRF' run, we find $86$ per cent of the H$_{2}$ mass in the central $4 \, \rm{kpc}$ of the disc lies in pixels with $I_{\rm{CO}} < 0.25 \, \rm{K} \, \rm{km} \, \rm{s}^{-1}$ if we use non-equilibrium abundances, compared to $59$ per cent if we assume chemical equilibrium. 
\end{enumerate}

To summarise, we have demonstrated that metallicity and UV radiation affect the global properties of galaxies, such as their star formation rates, and the observable signatures of individual chemical species. In contrast, non-equilibrium chemistry and cooling generally do not strongly affect the global properties of galaxies, although they do affect the observable diagnostics, particularly in molecular gas. 

To investigate how sensitive these results are to the resolution of the simulations, we also repeated the ref simulation twice, with a mass resolution a factor of four higher and lower than our fiducial resolution of $750 \, \rm{M}_{\odot}$. We find that our results are mostly unaffected by resolution over the range that we test here. In particular, the star formation rates (Fig.~\ref{SFH}) are similar at the three different resolutions, although the low resolution run shows a smoother star formation history with no strong bursts. The mass loading factors of the outflows (Fig.~\ref{mass_loading_fig}) are also similar at different resolutions. The mass of outflowing molecular hydrogen (Fig.~\ref{H2_outflow_fig}) is similar at our fiducial and high resolutions, although the low resolution run shows less difference between non-equilibrium and equilibrium in the molecular outflows. 

However, there are several important caveats that we need to highlight that may mean that we have underestimated the importance of non-equilibrium chemistry in these simulations. Firstly, we simulate isolated galaxies that do not include cosmological processes such as galaxy mergers and accretion of gas onto the galaxy. We might expect that such processes could enhance the non-equilibrium effects. For example, during and immediately after a merger, the ISM will evolve rapidly as it re-distributes its gas in the temperature-density plane. This could potentially drive chemical abundances further out of equilibrium. To investigate the importance of such processes for the chemistry, we could repeat this study using cosmological zoomed simulations of individual galaxy haloes. 

Secondly, turbulence in the ISM can also drive chemical abundances out of equilibrium. For example, \citet{gray15} recently presented a series of high-resolution simulations of the turbulent ISM. They showed that the steady-state ion abundances at the end of their simulations can be out of equilibrium by several orders of magnitude at high Mach numbers. If, as expected, we do not fully resolve such small-scale turbulence in our simulations, then we are likely to underestimate the importance of non-equilibrium chemistry. 

Thirdly, chemical abundances can also be driven out of equilibrium by a fluctuating radiation field. For example, \citet{oppenheimer13b} demonstrated that the abundances of metal ions in the circumgalactic medium can be affected by the presence of an AGN even after the AGN has turned off, as it takes a long time for the ions to recombine. In our simulations, we apply a local prescription for self-shielding of the radiation field by gas and dust, which does vary with position and time. However, we apply the self-shielding to a constant, uniform radiation field. In reality, the UV radiation from young stars will fluctuate as new stars are born, existing stars age and gas particles move in relation to the stars. This may drive additional non-equilibrium effects that we do not capture in our current simulations. 

Furthermore, since we have shown that UV radiation affects the global properties of galaxies, the inclusion of a fluctuating radiation field will also influence the galaxy directly, in addition to any non-equilibrium effects that it may drive. Likewise, it will also be important to include fluctuations in metallicity due to chemical enrichment from stars, which are not included in the fixed-metallicity models that we present here. 

\section*{Acknowledgments}

We are very grateful to Volker Springel for sharing \textsc{gadget}3 and his inital conditions code, and to Claudio Dalla Vecchia for allowing us to use \textsc{anarchy}. We thank everyone who has contributed to the development of the \textsc{eagle} code, which our simulations were partly based on, and we thank Rob Crain, Benjamin Oppenheimer, Simon Glover and Ewine van Dishoeck for useful discussions. We gratefully acknowledge support from the European Research Council under the European Union's Seventh Framework Programme (FP7/2007-2013) / ERC Grant agreement 278594-GasAroundGalaxies. This work used the DiRAC Data Centric system at Durham University, operated by the Institute for Computational Cosmology on behalf of the STFC DiRAC HPC Facility (www.dirac.ac.uk). This equipment was funded by BIS National E-infrastructure capital grant ST/K00042X/1, STFC capital grant ST/H008519/1, and STFC DiRAC Operations grant ST/K003267/1 and Durham University. DiRAC is part of the National E-Infrastructure. This work also used computer resources provided by the Gauss Centre for Supercomputing/Leibniz Supercomputing Centre under grant:pr83le. We further acknowledge PRACE for awarding us access to resource Supermuc based in Germany at LRZ Garching (proposal number 2013091919).

{}

\label{lastpage}

\end{document}